\documentclass[sigconf, screen]{acmart}
\renewcommand\footnotetextcopyrightpermission[1]{}

\usepackage{mathtools}
\usepackage{amsmath,amsfonts, amsthm}
\usepackage{graphicx,subcaption}    
\usepackage{textcomp}
\usepackage{xcolor}
\usepackage{xspace}
\usepackage{comment}
\usepackage{hyperref}
\usepackage{multicol}
\usepackage{multirow}
\usepackage{booktabs} %
\usepackage{algorithm}
\usepackage{listings}
\usepackage{centernot} %
\usepackage[noend]{algpseudocode}
\usepackage{adjustbox}
\usepackage{longfbox}
\usepackage{appendix}
\usepackage{placeins}
\usepackage{pdfpages}

\newtheorem{theorem}{Theorem}

\DeclarePairedDelimiter{\l1norm}{\lVert}{\rVert_1}

\DeclareMathOperator*{\argmax}{arg\,max}
\DeclareMathOperator*{\argmin}{arg\,min}

\usepackage[]{graphicx}
\graphicspath{{./figures/}}
\DeclareGraphicsExtensions{.pdf}

\renewcommand{\sectionautorefname}{\S\kern-2pt}

\newcommand{\sys}{FOX\xspace}

\newcommand{\numprograms}{38\xspace}
\newcommand{\numstandalone}{15\xspace} %
\newcommand{\numfuzzbench}{23\xspace} %
\newcommand{\numcpuyears}{6\xspace} %
\newcommand{\maxstandalone}{26.45\%\xspace} %
\newcommand{\maxfuzzbench}{6.59\%\xspace} %
\newcommand{\avgfuzzbench}{3.50\%\xspace} %
\newcommand{\maxmagma}{16.16\%\xspace} %
\newcommand{\totalbugs}{20\xspace} %
\newcommand{\aflpp}{AFLPP\xspace}
\newcommand{\cmplogdict}{AFLPP+CD\xspace}
\newcommand{\cmplog}{AFLPP+C\xspace}
\newcommand{\sysched}{\sys-SCHED\xspace}
\newcommand{\sysbase}{\sys-BASE\xspace}
\newcommand{\sysdict}{FOX+D\xspace}
\newcommand{\vx}{{\bf x}}

\begin{document}

\title{FOX: Coverage-guided Fuzzing as Online Stochastic Control}

\author{Dongdong She}
\authornote{Both authors contributed equally to the paper.}
\affiliation{%
\institution{Hong Kong University\\of Science and Technology}
\city{Hong Kong}
\country{China}}
\email{dongdong@cse.ust.hk}
\author{Adam Storek}
\authornotemark[1]
\affiliation{%
\institution{Columbia University}
\city{New York}
\state{NY}
\country{USA}}
\email{astorek@cs.columbia.edu}
\author{Yuchong Xie}
\affiliation{%
\institution{Hong Kong University \\of Science and Technology}
\city{Hong Kong}
\country{China}}
\email{yu3h0xie@gmail.com}
\author{Seoyoung Kweon}
\affiliation{%
\institution{Columbia University}
\city{New York}
\state{NY}
\country{USA}}
\email{sk4865@columbia.edu}
\author{Prashast Srivastava}
\affiliation{%
\institution{Columbia University}
\city{New York}
\state{NY}
\country{USA}}
\email{ps3400@columbia.edu}
\author{Suman Jana}
\affiliation{%
\institution{Columbia University}
\city{New York}
\state{NY}
\country{USA}}
\email{suman@cs.columbia.edu}

\begin{abstract}
Fuzzing is an effective technique for discovering software vulnerabilities by generating random test inputs and executing them against the target program. However, fuzzing large and complex programs remains challenging due to difficulties in uncovering deeply hidden vulnerabilities. This paper addresses the limitations of existing coverage-guided fuzzers, focusing on the scheduler and mutator components. Existing schedulers suffer from information sparsity and the inability to handle fine-grained feedback metrics. The mutators are agnostic of target program branches, leading to wasted computation and slower coverage exploration.

To overcome these issues, we propose an end-to-end online stochastic control formulation for coverage-guided fuzzing. Our approach incorporates a novel scheduler and custom mutator that can adapt to branch logic, maximizing aggregate edge coverage achieved over multiple stages. The scheduler utilizes fine-grained branch distance measures to identify frontier branches, where new coverage is likely to be achieved. The mutator leverages branch distance information to perform efficient and targeted seed mutations, leading to robust progress with minimal overhead.

We present \sys, a proof-of-concept implementation of our control-theoretic approach, and compare it to industry-standard coverage-guided fuzzers. \numcpuyears CPU-years of extensive evaluations on the FuzzBench dataset and complex real-world programs (a total of \numprograms test programs) demonstrate that \sys outperforms existing state-of-the-art fuzzers, achieving average coverage improvements up to \maxstandalone in real-world standalone programs and \maxfuzzbench in FuzzBench programs over the state-of-the-art AFL++. In addition, it uncovers \totalbugs unique bugs in popular real-world applications including eight that are previously unknown, showcasing real-world security impact. 
\end{abstract}

\maketitle

\section{Introduction}
Fuzzing is a popular technique for discovering software vulnerabilities by generating random test inputs and executing them against the target program~\cite{She2022EffectiveSS, Aschermann2019, bohme2020boosting, fioraldi_afl_2020}. While it has been successful in detecting security vulnerabilities in real-world programs~\cite{syzbot,ossfuzz}, fuzzing large and complex programs remains challenging due to difficulties in uncovering deeply hidden vulnerabilities.

This paper focuses on coverage-guided fuzzers, the prevailing approach to fuzzing, aiming to maximize edge coverage within a given time budget. These fuzzers maintain a list of seed inputs and select inputs for further mutation at each stage. They consist of two main components: a scheduler (choosing inputs for mutation) and mutators (modifying the chosen input). The goal is to generate inputs that explore new edges for better coverage. Most existing fuzzers use randomized mutations to adapt to different branches in the target program. The effectiveness of a fuzzer, therefore, depends on two factors: (i) the mutator's likelihood to generate new inputs achieving new coverage given a specific seed input, and (ii) the scheduler's ability to identify seeds that, when mutated, are likely to trigger new edges.

\vspace{0.2cm}
\noindent
{\bf Limitations of Existing Approaches.} We identify the following main drawbacks in the existing design of scheduler and mutator components for coverage-guided fuzzing. Firstly, the schedulers use coarse-grained feedback to select candidates for further mutation. They
rely on seeds that have previously resulted in coverage gain when mutated. However, this approach suffers from serious sparsity of information, as coverage-increasing inputs become increasingly rare as the fuzzing campaign progresses~\cite{nagy2019full}.
Consequently, the scheduler often degenerates into a round-robin approach.
Attempting to use finer-grained data-flow-guided feedback metrics~\cite{mantovani2022fuzzing, herrera2022dataflow} to address this issue can easily lead to an explosion in the seed corpus of the fuzzing scheduler~\cite{wang2019sensitive}. Secondly, the current mutators are agnostic of the target program branches. They perform random mutations independently of the branch logic, with the hope of increasing coverage. As a result, these existing mutators waste computation while attempting to generate coverage-increasing mutations, leading to slower coverage exploration. Techniques using taint tracking to customize mutation operations for different branches tend to incur prohibitively high overhead~\cite{wang2010taintscope}. Furthermore, despite the shared overarching goal of achieving new coverage, current fuzzers often treat the scheduler and mutator as separate entities with distinct objectives and little information exchange.

\begin{figure}[!]
\centering
\includegraphics[scale=0.30]{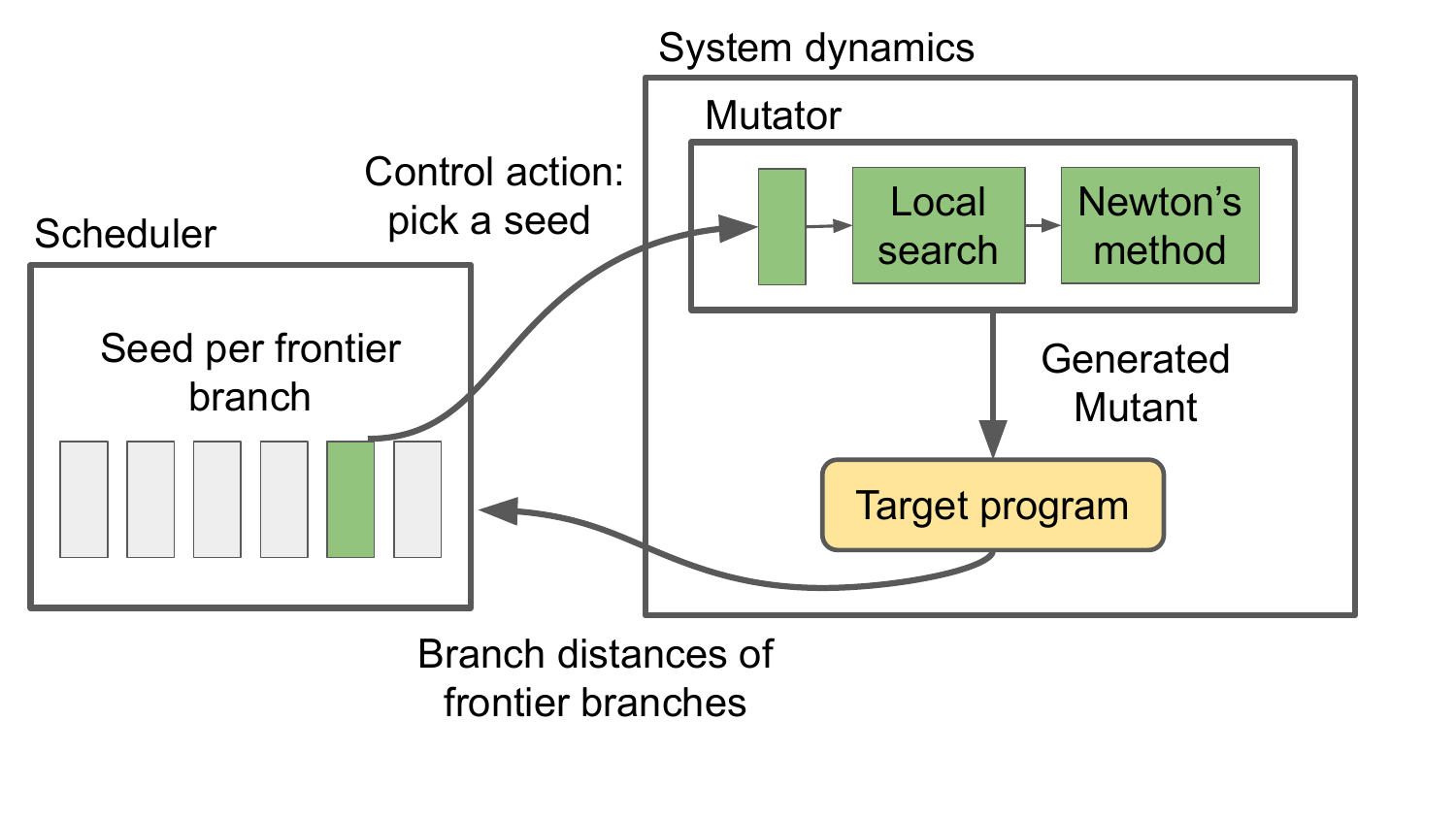}
\caption{\textbf{Workflow of \sys}}
\label{fig:flowchart}
\end{figure}

\vspace{0.1cm}
\noindent
{\bf Our Approach.} In this paper, we tackle these issues by presenting an end-to-end online stochastic control formulation for coverage-guided fuzzing, which encompasses both the scheduler and mutator components. In this framework, the stochastic mutator and target program represent the dynamics of the system, where the scheduler makes probabilistic online control decisions about which seed to mutate from the corpus, representing the fuzzer's state. Each scheduling step constitutes a stage of this control process, and our objective is to maximize the sum of expected coverage gain across multiple stages subject to a time budget constraint. To solve this problem, we introduce a novel scheduler and mutator that can efficiently adapt to branch logic, integrating them into a comprehensive control framework that can benefit from both the scheduler's multi-seed view and the mutator's seed- and branch-specific behavior. The workflow of \sys is shown in~\autoref{fig:flowchart}.

\vspace{0.1cm}
\noindent
{\bf Scheduler.} To address the lack of meaningful information for the scheduler when mutators fail to achieve new coverage, we use fine-grained branch distance measures (see \autoref{subsec: decomp}), indicating how close the current seed is to \emph{flipping} a branch. Specifically, \emph{flipping} indicates generating and executing a seed that can exercise alternative outgoing control-flow branches from the parent node of a previously seen branch. To avoid state space explosion while leveraging finer-grained feedback metrics, we apply the fine-grained feedback measure only to \emph{frontier branches} — branches that have at least one unexplored outgoing control-flow edge from their corresponding parent node in the control-flow graph. Thus, our control problem is simplified to maximizing coverage by flipping frontier branches at each stage (see Section~\ref{sec:theory} for a formal definition).

Our approach estimates the potential for new coverage of a seed based on the potential for a branch distance decrease at frontier nodes. New coverage implies a decrease in branch distance, though a branch distance decrease does not necessarily imply new coverage. Since branch distance decreases are much more frequent, we use them as a reliable proxy and an upper bound on the probability of achieving new coverage. The scheduler keeps track of the branch distance decreases, estimating the probability of a seed, when mutated, to flip a frontier branch with minimal additional overhead. We present a greedy online scheduling algorithm that leverages branch-distance-based probabilistic estimates of expected coverage gain to make provably optimal decisions among all possible scheduling algorithms with access to the same estimates of the probability of achieving new coverage.

\vspace{0.1cm}
\noindent
{\bf Mutator.} In order to overcome the limitations of randomized mutators in adapting to branch logic, we present a new mutator that utilizes branch distance information for each frontier branch. Our mutator has two components: local search and Newton's method. First, we use local search to identify new seeds reaching each frontier branch and efficiently learn an approximate linear lower bound of the branch function. Next, we generate new seeds based on Newton's method of root finding, with a high probability of flipping the target frontier branch. This approach leads to fast and robust progress for a wide range of branch distance functions, while maintaining minimal overhead compared to more expensive techniques like byte-level taint inference~\cite{Aschermann2019, gan2020greyone, patafuzz}.

Our design not only results in coverage gains but also provides actionable fine-grained, frontier-branch-specific feedback for fuzzing developers and users to debug and optimize their setup, going beyond the current work on fuzzing interpretability~\cite{introspector}.
For instance, our scheduler can estimate the different sources of difficulty in flipping frontier branches, including the error in the linear approximation of a branch and the rate of reachable samples to a branch.
\noindent
{\bf Result Summary.} To evaluate the effectiveness of our control-theoretic approach to fuzzing, we implement \sys and compare it against the leading state-of-the-art fuzzers: AFL++, AFL++ with cmplog, AFL++ with cmplog and dictonary.  We perform an extensive evaluation involving over \numcpuyears CPU-years of computation of the coverage achieved by \sys and the other evaluation candidates on a set of \numprograms programs (\numfuzzbench from the FuzzBench dataset~\cite{metzman2021fuzzbench} as well as \numstandalone from a manually curated dataset of complex real-world programs). \sys outperforms all existing state-of-the-art fuzzers on standalone programs, achieving average improvements as high as \maxstandalone compared to AFL++, $16.98\%$ compared to AFL++ with cmplog, and $12.90\%$ compared to AFL++ with cmplog and dictionary. The same performance trend is reflected in the FuzzBench dataset where we see \sys gain a coverage improvement up to \maxfuzzbench over AFL++ including an average performance improvement of \avgfuzzbench when compared across all fuzzers. In addition, we evaluated the bug discovery capabilities of \sys, comparing its performance against the other candidate fuzzers on Magma dataset \cite{hazimeh2020magma} with ground-truth bugs as well as its ability to find bugs in the wild for real-world applications. On the Magma dataset, \sys uncovers up to \maxmagma more ground-truth bugs compared to other fuzzers. Finally, \sys also uncovered $20$ unique bugs as part of its in-the-wild evaluation of real-world programs, of which eight were previously unknown and have been responsibly disclosed to the affected vendors. 

In summary, we make the following contributions:
\begin{itemize}
    \item Formulating fuzzing as an online stochastic control problem, presenting a unified framework for reasoning about the scheduler and mutator components in tandem.
    \item Performing coverage exploration by scheduling frontier branches, drastically reducing the scheduler's control space while employing finer-grained feedback in the form of branch distance.
    \item Designing branch-aware mutators using optimization-\\oriented strategies to gain new coverage and provide interpretable feedback about frontier branches' characteristics.
    \item Introducing \sys as a proof-of-concept of our stochastic-control-guided approach, outperforming existing state-of-the-art fuzzers (AFL++, AFL++ cmplog, AFL++ cmplog with dictionary) on FuzzBench as well as a set of complex real-world programs, achieving coverage improvements up to \maxstandalone.
    \item Releasing \sys at \url{https://github.com/FOX-Fuzz/FOX} to foster further research and collaboration within the research community. 
\end{itemize}

\section{Theory}
\label{sec:theory}
\subsection{Problem Definition}
Coverage-guided mutation-based fuzzing can be defined as an online optimization problem aiming to maximize the edge coverage of a target program. Let us begin with an arbitrary target program denoted as $P$, which can take inputs up to length $M$ bits. A fuzzer maintains a seed corpus of inputs and iteratively mutates these seeds to generate new inputs. These mutated inputs are then executed by the target program to determine the coverage achieved on program edges.

At each fuzzing iteration $i$, the fuzzer selects a seed $S_i[u_i]$ from the current seed corpus $S_i$, where $u_i$ represents the index of the chosen seed. Next, the fuzzer applies the mutator $mut(\cdot)$ to $S_i[u_i]$, generating a new input ${\bf x}$. The target program is then executed with this mutated input, and the achieved edge coverage increase is calculated using the function $cov({\bf x}, S_i)$ over the mutated input ${\bf x}$ and the seed corpus $S_i$.

Based on the coverage outcome, the fuzzer decides whether/how to update the seed corpus $S_i$ to  $S_{i+1}$. Typically, popular fuzzers add ${\bf x}$ to $S_i$ if it leads to a coverage increase.

The objective is to maximize the accumulated coverage across $K$ stages. This can be formally represented as follows:
\begin{equation}
\textit{Maximize } \sum_{i=1}^K \left[cov(mut(S_i[u_i]), S_{i})\right]
\label{eq: general fuzzing optimization objective}
\end{equation}

Here, $S_i[u_i]$ denotes the seed selected for mutation at stage $i$, and $cov(mut(S_i[u_i]), S_{i})$ represents the coverage achieved after executing the mutated input on the target program.

The optimal strategy for the fuzzer to achieve the highest coverage for a given target program is to find the sequence of indices $u_i$ into the seed corpus $S_i$ that maximizes the objective function representing the accumulated coverage across $K$ stages. This problem is an online optimization problem~\cite{hazan2016introduction} because the fuzzer can only observe the output of the $cov$ function after applying the mutation operation at each stage sequentially. In an online optimization setting, the decision-maker (in this case, the fuzzer) must make choices at each stage without complete knowledge of the objective function's values for future stages. The objective of the fuzzer is to maximize the cumulative coverage over the sequence of stages, leading to an online optimization problem where the optimal sequence of seed selections and mutations needs to be found to achieve the highest coverage across the entire fuzzing process.

\subsection{Fuzzing as Online Stochastic Control \label{subsec: fuzzing as online stochastic control}} 
One approach to addressing the online optimization problem described above (Equation~\ref{eq: general fuzzing optimization objective}) is to formulate it as an optimal control problem. This allows us to ground the development of fuzzing algorithms on the rich theory of optimal control. In an optimal control problem, the main goal is to find an optimal control strategy given a state space and control space. In our fuzzing scenario, the state space represents the state of the fuzzer at each stage, which is the current seed corpus $S_i$, and the control space corresponds to the choices the fuzzer makes, such as selecting a seed for mutation.

However, the dynamics of the target program are not fully known in advance due to the online nature of the problem. The system dynamics are revealed locally based on the actions selected by the fuzzer's mutation and control process. In essence, the fuzzer makes decisions at each stage based on the current seed corpus $S_i$, and then the program's behavior and response become apparent when the chosen mutated input is executed.

Moreover, most existing fuzzers employ randomized mutations to adaptively explore diverse program behaviors and various branch types. Since the input space of a fuzzer is often vast and complex, a randomized mutation approach avoids introducing specific biases toward particular inputs. Instead, it allows the fuzzer to explore a wide range of possibilities.

Due to both the mutator performing randomized mutations and the uncertainty in the target program's behavior, the system dynamic is best represented as a stochastic process. As a result, the problem at hand is considered a stochastic optimal control problem. In this context, the fuzzer aims to find an optimal mutation strategy that maximizes coverage despite the stochastic nature of the system dynamics. This probabilistic optimization approach allows the fuzzer to efficiently explore the input space and achieve effective generalization, ultimately improving coverage across diverse programs and input structures.

Formulating fuzzing as a stochastic control problem involves five key components as described below:

\noindent
\textbf{State Space.} The state space encompasses all possible configurations or states of the system at any given stage. In the context of fuzzing, it includes information about the current seed corpus at stage $i$, denoted as $S_i$.

\noindent
\textbf{Control Space.} The control space encompasses all the possible decisions that can influence the system's behavior. In the context of fuzzing, it refers to the choice of which seed, denoted with index $u_i$ in the seed corpus $S_i$ ($1 \leq u_i \leq |S_i|$), to select at each stage.

\noindent
\textbf{Fuzzer Dynamics.} The dynamics summarize how the system evolves over stages based on the chosen actions and the current state. In fuzzing, it is characterized by three main steps in the mutation stage:

First, the fuzzer takes the seed corpus $S_i$ and selects a seed $S_i[u_i]$ based on the control strategy. It then generates a new input ${\bf x}$ by applying the mutator $mut(\cdot)$ to the selected seed $S_i[u_i]$. Next, the fuzzer executes the generated mutant on the target program, resulting in a coverage value denoted by $g_i$. This coverage value represents the extent of coverage achieved for the specific mutated input. Finally, the fuzzer updates the seed corpus (i.e., state) $S_i$ to create a modified seed corpus $S_{i+1}$ based on the coverage result. This step involves incorporating the new input and its corresponding coverage into the seed corpus for the subsequent stages.

As mentioned earlier, the dynamics of the fuzzing process are stochastic due to the random nature of the mutator. Formally, we define the dynamics steps below:
\begin{equation}
\begin{gathered}
{\bf x} \leftarrow mut(S_i[u_i]) \\
g_i = \mathbb{E}\left[cov({\bf x}, S_i)\right] \\
S_{i+1} \leftarrow update(S_i, {\bf x}, g_i)
\end{gathered}
\end{equation}

\noindent
\textbf{Objective Function.} The objective function defines the goal or desired outcome of the system. In the context of fuzzing, it aims to maximize the expected total coverage gain across $K$ stages:
\begin{equation}
\textit{Maximize } \sum_{i=1}^K g_i
\label{eq: general fuzzing as stochastic control}
\end{equation}

\noindent
\textbf{Constraints.} These represent restrictions that need to be satisfied by the system. Fuzzing is typically subject to resource constraints like that the total execution time (i.e., the sum of execution time for each stage $t_i$) must not go over the total time budget $T$  of the fuzzing campaign:
\begin{equation}
\sum_{i=1}^K t_{i} \leq T
\label{eq: time constraint}
\end{equation}

\subsection{Decomposing $cov({\bf x}, S_i)$ \label{subsec: decomp}}
The coverage function depends on both the mutated input ${\bf x}$ and the current set of seen edges based on $S_i$. Formally, $cov({\bf x}, S_i)$ can be modeled as a random variable that counts how many unseen (i.e., not reached before by any input) edges at stage $i$ have been flipped by the randomized mutation operation performed on ${\bf x}$. To achieve this, we use indicator functions $cov_b({\bf x}, S_i)$, each corresponding to an unseen branch $b \in B_i$, where $B_i$ represents the set of unseen branches at stage $i$. These indicator functions are modeled as random variables taking values of $0$ or $1$, indicating whether the unseen branch $b$ is flipped or not. Utilizing the linearity of expectation, we can express the expectation of the $cov$ function in the following manner:
\begin{equation}
g_i = \sum_{b \in B_i} \mathbb{E}\left[cov_b({\bf x}, S_i)\right]
\end{equation}

Most modern fuzzers employ customized randomized mutators, denoted as $mut(\cdot)$, which essentially generate new inputs drawn from an unknown distribution specific to the fuzzer. In our analysis, we assume that the inputs generated by the mutator have a fixed, yet unknown, probability of flipping a given unseen branch $b$ (denoted as $\Pr(b \text{ flips} \mid mut(S_i[u_i]))$). Furthermore, different program inputs generated by $mut(\cdot)$ for different seeds are independently distributed. Popular existing fuzzers like AFL++ satisfy these assumptions. Mutators such as \texttt{havoc} randomly alter inputs without targeting specific branches, thus maintaining an equal chance of flipping any branch for a given seed. Likewise, the creation of one input does not influence the creation of another unless a branch is flipped, resulting in coverage gain.

As a result of these assumptions, we can model the coverage of a specific branch $b$, denoted as $cov_b({\bf x}, S_i)$, where ${\bf x} \leftarrow mut(S_i[u_i])$, as a random variable $X$ following a Bernoulli distribution with the probability of success given by $\Pr(b \text{ flips} \mid mut(S_i[u_i]))$.:

\begin{equation}
g_i = \sum_{b \in B_i} \Pr(b \text{ flips} \mid mut(S_i[u_i]))
\end{equation}

To maximize the expected coverage gain in stage $i$, we need a good estimate for $\Pr(b \text{ flips} \mid mut(S_i[u_i]))$. However, the conventional approach of observing the frequency with which $mut(S_i[u_i])$ flips the branch does not work for our task. This is because using this approach, the probability for not-yet-flipped branches will be zero, whereas this probability is irrelevant to us once we have flipped a branch.

To address this issue, we utilize the concept of frontier nodes which is closely related to the concept of horizon nodes introduced by She et al.~\cite{She2022EffectiveSS} in the context of graph-centrality-based seed scheduling. Consider a control flow graph (CFG) as $G=(N, E)$, where $N$ denotes a set of basic blocks and $E$ denotes control flow transitions between these blocks. Given a seed corpus $S$, we can classify $N$ into a set of unvisited nodes $U$ and another set of visited nodes $V$ based on the code coverage information ($S.cov$ indicates whether a node $n$ has been reached by the seed corpus $S$):
\begin{equation}
\begin{gathered}
V=\{n | n \in N, S.cov(n) = 1\} \\
U=\{n | n \in N, S.cov(n) = 0\}
\label{eq: cfg visited unvisited}
\end{gathered}
\end{equation}
Unlike She et al.~\cite{She2022EffectiveSS} who identify horizon nodes as a set of unvisited nodes lying at the boundary between visited and unvisited code regions, we define frontier nodes as a set of visited nodes that dominate all the unvisited nodes:
\begin{equation}
F=\{v | (v, u) \in E, v \in V, u \in U\}
\label{eq: frontier nodes set}
\end{equation}
For the purposes of using frontier nodes to understand coverage, we focus solely on the subset of frontier nodes containing control instructions that result in conditional jumps. Therefore, in our setting, each frontier node is associated with a conditional jump, i.e., a  \textbf{frontier branch}. Frontier branches at stage $i$ are denoted by $FB_i$.
We assume that each frontier branch $b$ at stage $i$ involves the evaluation of a predicate $Q_b$. The predicate evaluates to true or false based on a relation $f_b({\bf x}) \mathrel{R_b} 0$, where $\mathrel{R_b}$ represents the condition type ($<$, $\leq$, $>$, $\geq$, $=$, $\neq$) and ${\bf x}$ indicates a test input. \textbf{Flipping a branch} indicates finding ${\bf x}'$ such that $Q_b(f_b({\bf x}')) \neq Q_b(f_b({\bf x}))$. As part of fuzzing instrumentation added to the target program, we assume that our fuzzer has access to the 2-tuple $(Q_b(f_b(\bf x)), R_b)$ for each frontier branch $b \in FB_i$. With this information, we define branch distance function $\delta_b$ as a linear or piece-wise linear function of $f_b({\bf x})$ for each frontier branch $b$ as given in~\autoref{tab:piecewise}.

Consider a simple example: a branch $b$ \texttt{if ($x \leq 15$)}, meaning $f_b(x) = x - 15$, with a reaching input $x = 5$. The branch is frontier because we have not found an input that exceeds $15$. We aim to define a linear branch distance function that outputs how close $x$ is to exceeding $15$. For an input $x = 5$, $Q_b(f_b(x)) = true$ (i.e., $x \leq 15$ is true) and $R_b =$ $\leq$. Consequently, according to~\autoref{tab:piecewise}, the employed function is $1-f_b(x)$ or $16 - x$. Finding the root input of $16 - x$, i.e. $x=16$, will flip the frontier branch $b$.

\begin{table}[!htbp]
\caption{\label{table: br dist table} Branch distance function $\delta_b$ for a frontier branch $b \in FB_i$ based on the 2-tuple $(Q_b(f_b({\bf x})), R_b)$. 2-tuples mapping to the same branch distance function are grouped together.}
\label{tab:piecewise}
\centering
\begin{tabular}{c c||c}
\toprule
$Q_b(f_b({\bf x}))$ & $R_b$ & $\delta_b$({\bf x}) \\ \hline
 false & $<$  & \multirow{2}*{$f_b({\bf x}) - 1$} \\
 true & $\geq$  & \\ \hline
 false & $\leq$ & \multirow{2}*{$f_b({\bf x})$} \\
 true & $>$  & \\ \hline
 false & $>$ & \multirow{2}*{$1 - f_b({\bf x})$} \\
 true & $\leq$ & \\ \hline
 false & $\geq$ & \multirow{2}*{$- f_b({\bf x})$} \\
 true & $<$ & \\ \hline
 false & $=$ & \multirow{2}*{$|f_b({\bf x})|$} \\
 true & $\neq$ & \\ \hline
 false & $\neq$ & \multirow{2}*{$1 - |f_b({\bf x})|$} \\
 true & $=$ & \\
 \bottomrule
\end{tabular}

\end{table}

Due to the nature of existing mutators as discussed above, we further assume that the program inputs generated by the mutator have an unknown but fixed probability of decreasing branch distance $\delta_b$ for a given unseen branch $b$ ($\Pr(\delta_b \text{ decreases} \mid mut(S_i[u_i]))$). Therefore, we can model the event $\delta_b({\bf x}) \text{ decreases}, {\bf x} \leftarrow mut(S_i[u_i])$ as a random variable $Y \sim Bernoulli(\Pr(\delta_b \text{ decreases} \mid mut(S_i[u_i])))$. While it is evident that $X$ implies $Y$ (every flip always involves a branch distance decrease), it is not necessarily true that $Y$ implies $X$ (not every branch distance decrease involves a flip). Hence, we find that: $\Pr(b \text{ flips} \mid mut(S_i[u_i])) \leq \Pr(\delta_b \text{ decreases} \mid mut(S_i[u_i]))$. Therefore, optimizing based on $p'$ serves as an approximation to optimizing based on $p$, where $p'$ and $p$ are the probabilities of $\delta_b$ decreasing and $b$ flipping, respectively. 

Naturally, $p'$ cannot be used to reason about unseen branches, i.e., branches that have not been reached yet, since we lack branch distance $\delta_b$ information for them. Thus, we focus solely on frontier branches $FB_i$ — branches that we have reached but not yet flipped.
\begin{equation}
g_i \approx \sum_{b \in FB_i} \Pr(\delta_b \text{ decreases} \mid mut(S_i[u_i]))
\label{eq: br dist prob formulation}
\end{equation}

To optimize Equation \ref{eq: br dist prob formulation}, we use a two-pronged approach. First, we choose to schedule $S_i[u_i]$ for each stage $i$ such that it maximizes the above expression. Second, assuming that most frontier branches in a program are approximately linear in the neighborhood of each seed in $S_i$, we introduce a custom mutation operation such that the $\Pr(\delta_b \text{ decreases} \mid mut(S_i[u_i]))$ increases for these branches.

\subsection{Fuzzing Algorithm \label{subsec: scheduling}}
\subsubsection{Optimal-control-based Scheduler}
We solve the following optimization problem using a greedy approach.
\begin{equation}
Maximize \sum_{i=1}^{K}\sum_{b \in FB_i} \Pr(\delta_b \text{ decreases} \mid mut(S_i[u_i]))
\label{eq: optimization problem}
\end{equation}
To optimize each stage $i$ efficiently, we aim to select a seed $S_i[u_i]$ that maximizes the inner sum in \autoref{eq: optimization problem}. Given that many of these probabilities are very small (0 or close to 0), we can approximate the inner sum by taking the maximum term instead. This approximation is not only effective but also computationally inexpensive in a streaming setting.
\begin{equation}
Maximize \sum_{i=1}^{K} \max_{b \in FB_i} \Pr(\delta_b \text{ decreases} \mid mut(S_i[u_i]))
\label{eq: optimization problem simplified}
\end{equation}
We can tackle this problem in two steps: (a) for each branch $b$, find a seed $S_i[u_i]$ that maximizes $\Pr(\delta_b \text{ decreases} \mid \text{mut}(S_i[u_i]))$ and (b) select the frontier branch $b \in FB_i$ with the maximum $\Pr(\delta_b \text{ decreases} \mid \text{mut}(S_i[u_i]))$.  For step (a), we maintain a mapping $TS$ where each frontier branch $b$ at stage $i$ corresponds to a seed $TS_i[b].s$ (along with the corresponding branch distance $TS_i[b].d$) that achieves the lowest branch distance across all inputs reaching that frontier branch so far. Subsequently, scheduling can be simplified to step (b) as shown below.
\begin{equation}
\begin{gathered}
    u_{i+1} = TS_i[\argmax_{b \in FB_i} \Pr(\delta_b <TS_i[b].d \mid mut(TS_i[b].s)] \\
    TS_{i+1}[b].s = \argmin_{s \in S_{i+1}} \delta_b(s) \\ %
    TS_{i+1}[b].d = \min_{s \in S_{i+1}} \delta_b(s) %
\end{gathered}
\end{equation}

To estimate the branch distance decrease probability for a frontier branch, we measure the number of total executions of inputs that reach the frontier branch, denoted as $th(\cdot)$, and the total number of inputs that lower the global minimum branch distance for the frontier branch, denoted as $ph(\cdot)$. To incorporate the time constraint, we further refine the probability estimate by replacing the number of hits with time. Specifically, we track the total time spent executing inputs that reach a frontier branch $tt(\cdot)$ and the total time spent executing inputs lowering the branch distance for a frontier branch $pt(\cdot)$. Additionally, to prevent frequent scheduling of the same seeds, we introduce a discount factor $\lambda_b$ based on how many times the seed mapped to the frontier branch $b$ was scheduled. The final probability estimate is then given by:
\begin{equation}
\Pr(\delta_b \text{ decreases}) = \lambda_b \frac{pt(b)}{tt(b)}
\end{equation}

\begin{theorem}
Given a fixed branch flip probability before each stage $i$, a greedy schedule that chooses $u_{i+1}$ such that it maximizes the expected coverage gain at each stage $i$ of the problem described by \autoref{eq: general fuzzing as stochastic control} is optimal.
\label{thm: greedy schedule optimal}
\end{theorem}

A proof of the theorem can be found in \autoref{thm proof: greedy schedule optimal}.

\subsubsection{Optimizing Mutation \label{subsec: mutation}}
In our approach, the primary objective of the mutation stage is to enhance the likelihood of reducing branch distances for all frontier branches at any given stage. To achieve this, we begin by creating a locally correct linear under-approximator through local search, essentially a subgradient. If the slope of this approximator is non-zero (indicating a decreasing direction), we proceed with the Newton's method.

\vspace{0.1cm}
\noindent
{\bf Local Search.} The central idea is to construct a local linear approximation of the branch distance function, denoted as $\delta_b({\bf x})$, for a given branch $b$. This local approximation should act as a lower bound on $\delta_b$ within a neighborhood $N({\bf x})$ surrounding the point ${\bf x}$. The local neighborhood depends on the number of bytes modified by the mutator in stage $i$. Unlike aiming to minimize the average error over the points within the neighborhood, we aim to minimize the under-approximation error. This choice ensures that the Newton's method remains stable.

To achieve this, we seek a vector $g$ that satisfies the condition $\forall {\bf x}' \in N({\bf x}): g^T \cdot ({\bf x}' - {\bf x}) \leq \delta_b({\bf x}')-\delta_b({\bf x})$. To approximate $g$, we resort to a randomized local search, generating a fixed-size sample of program inputs ${\bf x}' \in N({\bf x})$ using $mut(\cdot)$. We estimate the value of \(g\) by selecting $g = \argmax_{g'} \l1norm{g'}$, $g' = (\delta_b({\bf x}') - \delta_b({\bf x})) \oslash ({\bf x}' - {\bf x})$, where $\oslash$ indicates element-wise division.

The success of local search hinges on finding the right balance between accuracy and speed. A large stack of havoc mutations leads to sampling over a vast neighborhood $N({\bf x})$, rendering an approximation imprecise. Consequently, by fine-tuning the mutation stack we can constrain the neighborhood in a manner that allows for relatively more precise approximations. Details pertaining to this fine-tuning of the mutator stack as implemented in \sys are discussed in~\autoref{subsec:mutimpl}.

\vspace{0.1cm}
\noindent{\bf Newton's Method.} If we obtain a valid $g$ (i.e., non-zero norm) in the local search, we utilize it to perform the Newton's method along the direction of $g$ and identify the point at which the branch flips. Considering that we want to find an ${\bf x}$ such that $\delta_b({\bf x})$ takes a value of 0 and flips the branch, we apply Newton's method to generate a new input ${\bf x} = {\bf x} - \delta_b({\bf x}) \oslash g$. If the underlying function $\delta_b({\bf x})$ is linear and the branch is feasible, this step will cause the branch to flip. However, if $\delta_b(\vx)$ is not linear but a well-behaved convex function, the step is still likely to decrease the branch distance \cite{Boyd_Vandenberghe_2004}.

\section{Implementation}\label{sec:implementation}
In this section, we present \sys, our proof-of-concept implementation for modeling fuzzing as an online stochastic control problem. Two essential primitives for \sys to implement its stochastic-control-guided strategy include: (i) identifying frontier branches and (ii) calculating branch distances. We first discuss how \sys manages these two primitives, outlining the strategies employed for efficient tracking. Subsequently, we discuss the implementation of scheduling and mutation strategies, as described formally in~\autoref{sec:theory}. Finally, we explore how \sys leverages the semantics of string comparisons to effectively flip frontier nodes involved in such comparisons.

\subsection{Frontier Branch Identification}

\sys uses an intra-procedural control-flow graph (CFG) in conjunction with coverage information obtained during a fuzzing campaign to identify frontier branches dynamically. \sys augments the AFL++ LLVM-based instrumentation pass to extract the intra-procedural CFG for each function and embeds its metadata in the binary.

During a fuzzing campaign, we dynamically mark nodes in the CFG as either visited or unvisited based on coverage information as defined in~\autoref{eq: cfg visited unvisited}. Additionally, we update the list of frontier nodes following~\autoref{eq: frontier nodes set}. We only focus on frontier branches, a subset of frontier nodes where each node contains control instructions resulting in conditional jumps as outlined in~\autoref{subsec: decomp}.

\subsection{Branch Distance Tracking}\label{subsec:branchdistance}
To facilitate the tracking of branch distances, we employ an LLVM pass to hook every branch instruction and obtain its corresponding branch distance. Specifically, we examine each conditional branch instruction and check if its condition is computed from a CMP instruction. If so, we insert a function immediately after the CMP instruction to intercept its two operands, \texttt{op1} and \texttt{op2}.

\textbf{Integer Comparison}. We compute the branch distance $\delta_b(\vx) = \texttt{op1} - \texttt{op2}$.

\textbf{String Comparison}. We calculate an array of byte-wise branch distances $op1[i] - op2[i], 1 \leq i \leq k$, where $k$ corresponds to $\max(\text{len}(str1), \text{len}(str2))$. We add zero-byte padding to ensure both strings have the same length if one is shorter than the other.

During the dynamic execution of instrumented programs, our hook functions compute the branch distances and save them into a shared memory accessible by \sys. In real-world programs, there can be many branch instructions, and invoking a hook function for every branch instruction would cause significant runtime overhead during dynamic execution. To reduce runtime overhead, we implement an adaptive switch for each hook function. The switch ensures that the hook function is invoked only for frontier branches, not for fully explored branches where all children nodes have been visited. This way, we only incur a relatively small runtime overhead while accurately computing branch distances for a small set of frontier branches.

\subsection{Scheduler}
Our seed scheduler is implemented according to \autoref{alg: scheduling}, following the theoretical description provided in~\autoref{subsec: scheduling}. We maintain a mapping of seeds that achieve the lowest branch distance for each frontier branch $TS$. This approach allows us to reason about frontier branches rather than individual seeds, mitigating the target explosion issue discussed by Mouret et al. \cite{Mouret2015IlluminatingSS}.

We also keep counters for each frontier branch's productive time $PT$ and total time $TT$ indexed by the branch's unique CFG node id. These counters, along with the mapping $TS$, get updated after each program input execution. Before scheduling the next seed for each stage $i$, we compute the set of unvisited nodes $U_i$ and the set of frontier branches $FB_i$ following the definitions in ~\autoref{subsec: decomp}.

To compute the probability, we define the seed decay factor $\lambda_b$ for each frontier branch $b$ as $SC[TS[b.id]]$, where $SC$ is an array of schedule counters indexed by the seed. Instead of directly computing $\Pr(\delta_b \textit{ decreases})$ for each frontier branch $b$, which might suffer from numerical underflows if $PT[b.id]$ is much smaller than $TT[b.id] \times \lambda_b$, we compute the log-probability, as it avoids underflow issues.

\begin{algorithm}[]
\caption{\sys scheduling algorithm for stage $i$} 
\label{alg:scheduling} 
\lstset{basicstyle=\ttfamily\footnotesize, breaklines=true}
\begin{tabular}{|lp{2.7in}|}\hline
\textbf{Input}:
    & \textbf{$PT$} $\leftarrow$ CFG node id to productive time mapping \\
    & \textbf{$TT$} $\leftarrow$ CFG node id to total time mapping \\
    & \textbf{$TS$} $\leftarrow$ CFG node id to lowest branch distance seed mapping \\
    & \textbf{$SC$} $\leftarrow$ Seed to number of scheduled count mapping \\
    
\hline
\end{tabular}
\begin{algorithmic}
\State
\State $b\_max = \emptyset$
\State $b\_max\_logprob = \infty$
\For{$f \in FB_i$}
    \State $b\_logprob = \log(PT[b]) - \log(TT[b]) - \log(SC[TS[b].s])$
    \If{$b\_max\_logprob < b\_logprob$}
        \State $b\_max = b$
        \State $b\_max\_logprob = b\_logprob$
    \EndIf
\EndFor
\State $u_i = TS[b\_max].s$ \Comment{\textcolor{purple}{Schedule $top\_seed$}}
\State $SC[u_i] \mathrel{+}= 1$
\end{algorithmic}
\label{alg: scheduling}
\end{algorithm}

\subsection{Mutator}\label{subsec:mutimpl}
Our mutator is implemented as shown in \autoref{alg:mutating}, following the theoretical description provided in~\autoref{subsec: mutation}. Given the scheduled seed $S_i[u_i]$, we perform a randomized local search by sampling $k=1024$ program inputs $\vx$ in the neighborhood $N(S_i[u_i])$. We empirically determined this to be a good trade-off between the number of local searches performed and Newton's method steps taken. We utilize AFL++ \texttt{havoc} mode stochastic mutator~\cite{havoc} for generating mutants, with the number of random perturbations of the seed reduced to keep them within the local neighborhood $N(S_i[u_i])$.

For each generated mutant $\vx$, we determine the subset of frontier branches $FB_{i, \vx} \subseteq FB_i$ it reaches and compute the subgradient $g_{\vx, b}$ for each frontier branch $b \in FB_{i, \vx}$. We only consider a mutant for the Newton's method when the L1 norm of $g_{\vx, b}$ is greater than the maximum subgradient value $G[b.id]$ encountered so far. We then apply Newton's method to derive new inputs $\vx^*$ for the reached frontier branches and execute them on the target program.

\begin{algorithm}[!]
\caption{\sys mutator} 
\label{alg:mutating} 
\lstset{basicstyle=\ttfamily\footnotesize, breaklines=true}
\begin{tabular}{|lp{2.7in}|}\hline
\textbf{Input}:
    & \textit{$\delta$} $\leftarrow$ branch distance function for each frontier branch $b$ \\
    & \textit{$S_i$} $\leftarrow$ Seed corpus \\
    & \textit{$u_i$} $\leftarrow$ Scheduled seed index \\
    & \textit{$k$} $\leftarrow$ Local search sample size \\
\hline
\end{tabular}
\begin{algorithmic}
\State \textcolor{purple}{/* Perform local search with sample size k */}
\State $G = empty\_map()$
\State $X = empty\_map()$
\For{$k \textit{ iterations}$}
    \State $\vx = mut(S_i[u_i])$
    \For{$f \in FB_{i, \vx}$}
        \State $g = \textsf{ComputeSubgradient}(\vx, S_i[u_i], \delta_b)$
        \If{$b.id \notin X \lor \l1norm{g} > \l1norm{G[b.id]}$}
            \State $G[b.id] = g$
            \State $X[b.id] = \vx$
        \EndIf
    \EndFor
\EndFor
\State \textcolor{purple}{/* Apply Newton method to the top input for each reached frontier branch */}
\For{$f \in FB_i$}
    \If{$b.id \in X.keys$}
        \State $\vx^* = \textsf{ApplyNewtonMethod}(X[b.id], S_i[u_i], \delta_b)$
    \EndIf
    \State $exec(\vx^*)$
\EndFor
\end{algorithmic}
\label{alg: mutations}
\end{algorithm}

For frontier branches that rely on string comparisons, we have taken a different approach. We treat string comparison functions like \texttt{strcmp} and \texttt{strncmp} as sequences of multiple single-byte integer comparisons and solve them together using Newton's method. In contrast to the standard local search, for these functions, we take an additional step to analyze the byte differences between the seed $S_i[u_i]$ and the mutant $\vx$ to investigate the effect of a branch distance $\delta_b$ change.

For each byte difference, we create a new program input by applying only a single byte difference to the seed and then execute this modified input to observe if it leads to a change in branch distance. If we detect a branch distance change, we identify the specified byte difference in the input as a hot (i.e., influential) byte, indicating that it directly influences a particular byte in the string comparison.

Our approach is designed to efficiently identify hot byte locations without the need to probe all byte locations, which is done in many previous works ~\cite{Aschermann2019,gan2020greyone,liang2022pata}. Once we find a hot byte location and know the length of the string being compared, we infer all the other hot byte locations by exploiting the fact that the hot bytes for a string comparison are typically adjacent.

\section{Evaluation}
In this section, we aim to answer the following research questions:
\begin{itemize}
\item \textbf{RQ1:} Can \sys enable testing code that was previously unreachable by state-of-the-art fuzzers? %
\item \textbf{RQ2:} Does \sys improve fuzzers' bug discovery capabilities? %
\item \textbf{RQ3:} How much control space reduction can \sys achieve compared with existing fuzzers?
\item \textbf{RQ4:} To what degree do the individual components of \sys contribute to its overall performance? %
\item \textbf{RQ5:} Are there characteristics that make a branch amenable to be solved by \sys?
\end{itemize}

We perform a thorough evaluation of \sys against AFL++ version 4.09c (AFLPP) \cite{fioraldi_afl_2020}, which is the latest and most performant version of a widely-recognized state-of-the-art fuzzer \cite{asprone2022comparing, liu2023sbft}, topping the December 2023 FuzzBench report \cite{aflpp_fuzzbench_report}. Additionally, we extend our comparison to AFL++ in \texttt{cmplog} mode (\cmplog), an industry implementation of REDQUEEN~\cite{Aschermann2019}. The \cmplog mode intercepts the operands of CMP instructions and applies tailored mutations. We specifically opt for this comparison because the mutation technique of \cmplog most closely aligns with our own mutator. The most potent setting of AFL++, widely embraced in industrial environments~\cite{fuzzbenchsetup} and academic fuzzing competitions~\cite{fuzzcomp}, is AFL++ with both \texttt{cmplog} and \texttt{dictionary} modes (\cmplogdict). The \texttt{dictionary} mode of AFL++ automatically generates a fuzzing dictionary comprising string constants and integer constants extracted from the tested program. We also incorporate dictionaries provided by FuzzBench. In addition to \sys alone, we evaluate \sys with \texttt{dictionary} (\sysdict) to showcase its performance in a fair comparison against \cmplogdict.

\noindent{\textbf{Benchmarks.}}
For our evaluation, we use all \numfuzzbench binaries from the FuzzBench dataset~\cite{metzman2021fuzzbench}. While FuzzBench is widely acknowledged as a standard in fuzzer evaluation \cite{liu2023sbft, nicolae2023neuzzpp, asprone2022comparing, bohme2022reliability}, we observe that it suffers from two significant challenges: many project harnesses are either very small (about a third of binaries have less than 20,000 edges, see \autoref{tab:programs}) or feature numerous test cases, leading to an early coverage saturation. This makes it challenging to differentiate the performance differences among fuzzers. Notably, in SBFT23, a FuzzBench-based competition, the mean coverage gain of the winner HasteFuzz over \cmplogdict was a modest 1.28\%~\cite{fuzzcomp_report}. To address these limitations and provide a more comprehensive assessment, we further evaluate \sys on \numstandalone real-world standalone programs. These programs are carefully selected to represent the diversity of our current software ecosystem, encompassing a broad range of functionalities based on prior fuzzing literature~\cite{jauernig_darwin:_2023, fishfuzz2023zheng, aifore2023shi, She2022EffectiveSS}. The details of these programs, along with their corresponding commit/version, are presented in Table~\ref{tab:programs}. Our results demonstrate that the improvements achieved by \sys generalize beyond FuzzBench.

\begin{table}[]
\centering
\caption{\textbf{Studied programs in our evaluation.}}
\label{tab:programs}
\setlength{\tabcolsep}{1.8pt}
\renewcommand{\arraystretch}{1.1}
\begin{tabular}{llr|llr}
\toprule
\textbf{Targets}  & \textbf{Version} & \textbf{\# Edge} & \textbf{Targets}  & \textbf{Version} & \textbf{\# Edge} \\
\midrule
\multicolumn{3}{c|}{\textbf{FuzzBench Targets}} &
systemd        & 07faa49 & 127,142 \\ \cline{1-3}
bloaty        & 52948c1 & 163,494 &
vorbis        & 84c0236 & 13,729 \\
curl          & a20f74a & 122,270 &
woff2         & 8109a2c & 19,615 \\
freetype & cd02d35 & 34,951 &
zlib          & d71dc66 & 4,640 \\ \cline{4-6}
harfbuzz & cb47dca & 78,203 &
\multicolumn{3}{c}{\textbf{Standalone Targets}} \\ \cline{4-6}
jsoncpp & 8190e06 & 9,826 &
bsdtar & libarchive-3.6.2 & 38,244 \\
lcms          & f0d9632 & 14,693 &
exiv2 & exiv2-0.28.0 & 122,543 \\
libjpeg-t & 3b19db4 & 17,789 &
ffmpeg & ffmpeg-6.1 & 741,421 \\
libpcap       & 17ff63e & 15,344 &
jasper & jasper-4.1.2 & 19,122 \\
libpng        & cd0ea2a &  9,038 &
nm-new     & binutils-2.34    & 54,848 \\
libxml2       & c7260a4 & 81,782 &
objdump    & binutils-2.34    & 80,582 \\
libxslt       & 180cdb8 & 61,727 &
pdftotext       & xpdf-4.04        & 53,683 \\
mbedtls       & 169d9e6 & 28,373 &
readelf    & binutils-2.34    & 32,249 \\
openh264      & 045aeac & 19,599 &
size       & binutils-2.34    & 54,348 \\
openssl & b0593c0 & 80,662 &
strip-new       & binutils-2.34    & 61,051 \\
openthread      & 2550699 & 50,574 &
tcpdump    & tcpdump-4.99.4   & 47,476 \\
proj4 & a7482d3 & 267,147 &
tiff2pdf   & libtiff-v4.5.0   & 21,006 \\
re2           & b025c6a & 15,179 &
tiff2ps    & libtiff-v4.5.0   & 17,861 \\
sqlite3       & c78cbf2       & 77,154 &
tiffcrop   & libtiff-v4.5.0   & 20,096 \\
stb           & 5736b15 &  7,661 &
xmllint & libxml2-2.9.14 & 85,412 \\
\bottomrule
\end{tabular}
\end{table}

\noindent{\textbf{Experimental Setup.}}
To answer \textbf{RQ1}, we evaluate the coverage over time achieved by all fuzzers. For \textbf{RQ2}, we measure the number of real bugs found in the standalone programs as well as the time taken by the fuzzers to uncover the ground truth bugs present in the MAGMA dataset~\cite{hazimeh2020magma}. For \textbf{RQ3}, we quantify the number of frontier branches that \sys explores and compare it to the number of seeds that AFL++, a conventional coverage-guided fuzzer, has to schedule. To answer \textbf{RQ4}, we perform an ablation study by comparing the coverage achieved by \sys against its variant with the optimized mutator turned off. Finally, for \textbf{RQ5}, we propose and leverage a set of quantitative metrics to interpret the performance of \sys.

We follow standard fuzzing evaluation practices~\cite{klees2018evaluating} and run 10 campaigns of 24 hours each for each of the fuzzer-program pairs, totaling over 5 CPU-years of evaluation. As part of our evaluation on the FuzzBench dataset, we use the seeds as recommended in the benchmark. For our standalone program evaluation, we provide one well-formed seed for each of the targets sourced from the AFL++ repository~\cite{standalone_seeds}. To ensure fairness in our comparison, each fuzzer is assigned a single core for each of their fuzzing campaigns. In addition to standard summary statistics like mean and standard deviation over our previously mentioned evaluation metrics, we also perform the Mann-Whitney U test to ensure the observed performance differences are statistically significant. Since this test is non-parametric and therefore does not make any assumptions about the underlying distribution, it has been widely used in the software testing literature for testing randomized algorithms~\cite{tests2011arcuri} and fuzzers~\cite{klees2018evaluating}.
All our experiments are conducted on 10 machines with Intel Xeon 2.00 GHz processors running Debian bullseye with 100 GB of RAM.
\begin{figure*}[!]
        \includegraphics[width=\linewidth]{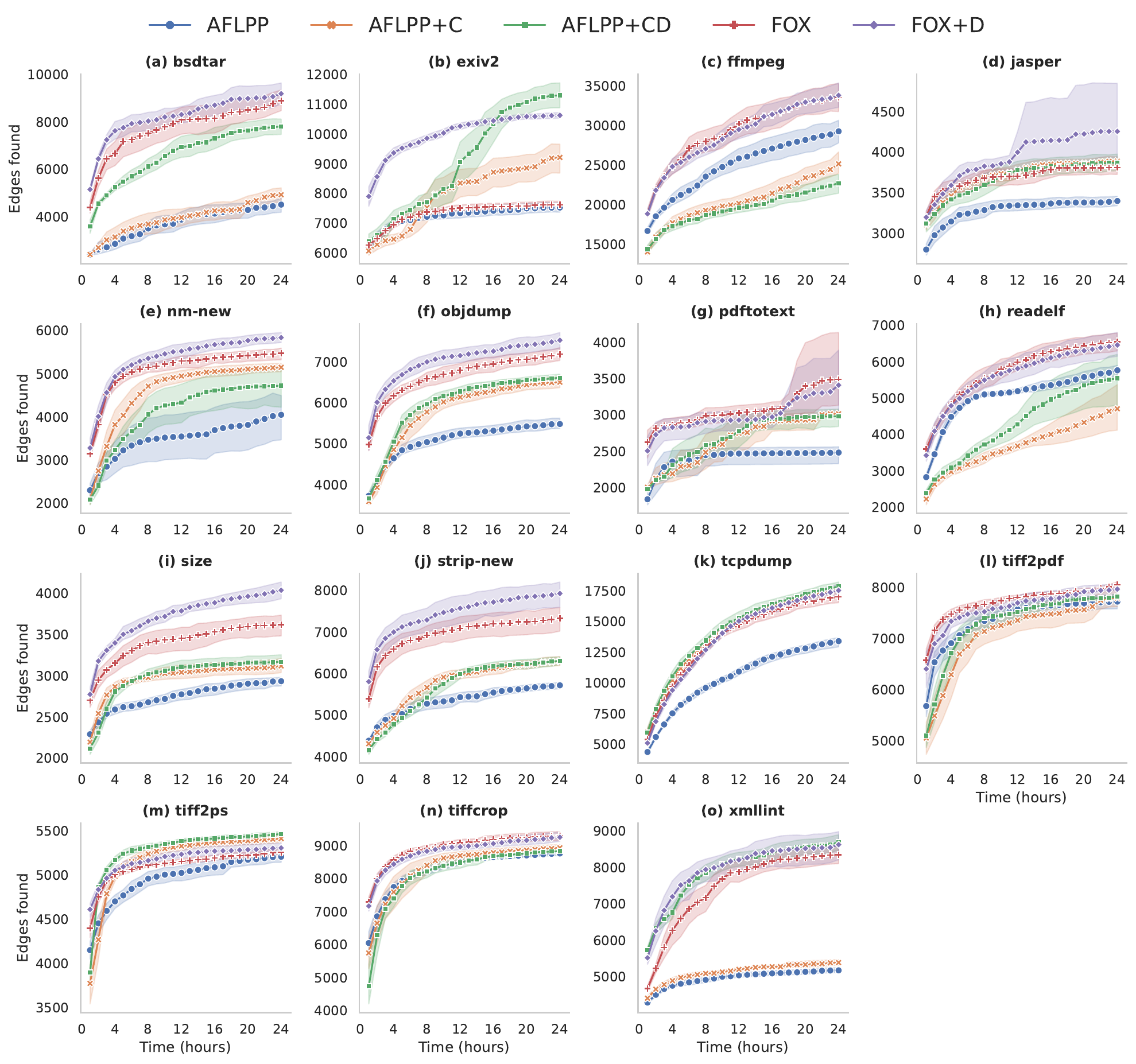}
	\caption{\small\textbf{The arithmetic mean edge coverage for \sys and \sysdict against three other fuzzers running for 24 hours over ten runs on the standalone programs. The error bars indicate one standard deviation.}}
	\label{fig:coverage2}
\end{figure*}

\subsection{RQ1: Code Coverage}\label{subsec:codecov}

To answer \textbf{RQ1}, we evaluate the code coverage achieved by \sys and compare it against the existing state-of-the-art fuzzers across \numfuzzbench FuzzBench and \numstandalone standalone targets on 10x24 hour campaigns. To quantify the coverage achieved, we leverage the coverage collection module of AFL++ that keeps track of the number of control-flow edges exercised during the run using its coverage bitmap. In addition, we compare the mean coverage growth over time along with their standard deviations to showcase the stability of our approach in achieving coverage saturation. The Mann-Whitney U test scores can be found in \autoref{eval:mann-whitney-u}. The total number of executions statistics can be found in \autoref{eval:throughput}. 

\noindent{\textbf{FuzzBench Targets.}}
Across FuzzBench targets (see results in \autoref{eval:fuzzbench}), \sys demonstrates superior performance compared to AFL++ on 18 programs and \cmplog on 13 programs, achieving significant mean coverage improvements. For instance, \texttt{freetype} shows up to a 39.97\% coverage increase over AFL++ and \texttt{harfbuzz} exhibits a 34.46\% improvement over \cmplog. Similarly, \sysdict outperforms \cmplogdict on 15 programs, with notable gains of up to 21.17\% on \texttt{harfbuzz}. Coverage progression over time, comparing \sys and \sysdict against other fuzzers, is detailed in \autoref{app:coverage}.

\sys performs comparably or with statistically insignificant differences to \cmplog on all remaining programs except for \texttt{lcms}, \texttt{openthread}, \texttt{proj4}, and \texttt{zlib}. The challenges in these targets stem from nested conditions and string comparisons with both operands as variables, requiring \sys to further adapt the Newton's method to handle these complex branch types. While our current prototype does not support them, it does not indicate a design limitation of \sys; we plan to extend its capabilities in future iterations (discussed further in~\autoref{sec:limitations}).

With the dictionary, addressing the bottleneck of variable string comparisons, \sysdict excels on larger FuzzBench programs with quality seed corpora like \texttt{harfbuzz} and \texttt{sqlite3}. Particularly noteworthy is its ability to manage control space more efficiently by focusing solely on frontier branches, reducing redundancy and maximizing time allocation. This effect is elaborated on in \autoref{eval:controlspace}. Even on relatively small programs, the Newton's method mutator allows \sys to flip branches that neither AFL++ nor \cmplog can flip. We further discuss this ability in \autoref{eval:introspection}.

\noindent{\textbf{Standalone Targets.}}
Comparison of mean coverage achieved by \sys against AFL++, \cmplog, and \cmplogdict is presented in \autoref{eval:standalone}. Across standalone programs, \sys exhibits improvements over AFL++ on all programs and surpasses \cmplog on 11 programs, achieving up to \maxstandalone more code coverage on average across the standalone target set. One of the targets where we see notable improvement is \texttt{bsdtar}, where \sys uncovers $97.25\%$ and $80.61\%$ more edges than AFL++ and \cmplog, respectively.   Similarly, \sysdict outperforms \cmplogdict on 11 programs, including a remarkable 49.04\% improvement on \texttt{ffmpeg}.

Given the larger size of our selected standalone programs compared to FuzzBench programs, tight integration of all fuzzer components becomes critical. Through comprehensive control space optimization, \sys effectively narrows the extensive control space further by prioritizing branches suitable for flipping by the Newton's method mutator, thereby enhancing coverage. In contrast, AFL++ relies solely on prior observed execution behavior to schedule seeds, lacking guidance on which seeds would be most effective in expanding coverage.

\vspace{0.1cm}
\begin{longfbox}
\textbf{Result 1:} 
\sys outperforms existing state-of-the-art fuzzers by achieving up to \maxstandalone more coverage on average across the standalone targets and up to \maxfuzzbench more coverage on average across the FuzzBench targets.
\end{longfbox}

\begin{table}[!ht]\caption{\small\textbf{Mean edge coverage of \sys and \sysdict against three fuzzers on \numfuzzbench FuzzBench programs for 24 hours over 10 runs. We mark the highest number in bold.}  }
    \centering
    \setlength{\tabcolsep}{2pt}
    \renewcommand{\arraystretch}{1.1}
    \label{eval:fuzzbench}
    \begin{tabular}{lrrr|rr}
    \toprule
        \textbf {Targets} & \textbf{\sys} & \textbf{\aflpp} & \textbf{\cmplog} & \textbf{\sysdict} & \textbf{\cmplogdict} \\ 
        \midrule
 bloaty        &    \textbf{8,646} &    8,507 &      8,627 &      8,514 &       \textbf{8,684} \\
 curl          &   \textbf{14,674} &   14,510 &     14,358 &     \textbf{15,811} &      15,359 \\
 freetype      &   \textbf{14,676} &   10,485 &     13,401 &     \textbf{15,536} &      13,812 \\
 harfbuzz      &   37,155 &   \textbf{37,276} &     27,633 &     \textbf{37,103} &      30,621 \\
 jsoncpp       &    1,341 &    1,341 &      \textbf{1,342} &      1,342 &       \textbf{1,343} \\
 lcms          &    1,224 &     728 &      \textbf{2,451} &      \textbf{1,976} &       1,882 \\
 libjpeg-turbo &    3,293 &    \textbf{3,299} &      3,297 &      3,283 &       \textbf{3,301} \\
 libpcap       &    \textbf{2,366} &      43 &      1,848 &      \textbf{3,074} &       2,690 \\
 libpng        &    \textbf{2,846} &    2,697 &      2,838 &      \textbf{2,848} &       2,836 \\
 libxml2       &   \textbf{18,757} &   18,623 &     18,616 &     \textbf{19,188} &      19,003 \\
 libxslt       &   \textbf{12,184} &   11,959 &     11,909 &     \textbf{12,424} &      12,358 \\
 mbedtls       &    \textbf{3,987} &    3,722 &      3,839 &      \textbf{4,037} &       3,834 \\
 openh264      &   \textbf{13,780} &   13,733 &     13,630 &     \textbf{13,826} &      13,662 \\
 openssl       &   \textbf{11,128} &   11,102 &     11,114 &     \textbf{11,127} &      10,902 \\
 openthread    &    4,827 &    4,686 &      \textbf{5,140} &      4,822 &       \textbf{5,041} \\
 proj4         &   26,603 &   26,703 &     \textbf{28,904} &     \textbf{32,995} &      32,494 \\
 re2           &    6,077 &    \textbf{6,190} &      6,105 &      6,085 &       \textbf{6,224} \\
 sqlite3       &   37,801 &   37,701 &     \textbf{37,953} &     \textbf{40,028} &      38,846 \\
 stb           &    \textbf{4,332} &    4,042 &      4,180 &      \textbf{4,440} &       4,249 \\
 systemd       &    3,836 &    3,781 &      \textbf{3,856} &      3,835 &       \textbf{3,857} \\
 vorbis        &    2,050 &    2,045 &      \textbf{2,056} &      2,054 &       \textbf{2,055} \\
 woff2         &    \textbf{2,674} &    2,458 &      2,550 &      \textbf{2,678} &       2,514 \\
 zlib          &     884 &     883 &       \textbf{917} &       879 &        \textbf{913} \\
        \midrule
        \multicolumn{2}{c|}{Mean gain}   & 6.59\% $\dagger$ & 0.95\% & --- &  2.97 \% \\
        \bottomrule
       \multicolumn{6}{l}{
       \begin{minipage}{8cm}
       \footnotesize $\dagger$ We omit \texttt{libpcap}, as it would unrealistically skew this statistic in favor of \sys.
       \end{minipage}
}
    \end{tabular}
\end{table}

\begin{table}[!ht]
\caption{\small\textbf{Mean edge coverage of \sys and \sysdict against three fuzzers on \numstandalone standalone programs over 10 fuzzing campaigns run for 24 hours each. We mark the highest number in bold.}  }
    \centering
    \setlength{\tabcolsep}{2pt}
    \renewcommand{\arraystretch}{1.1}
    \label{eval:standalone}
    \begin{tabular}{lrrr|rr}
    \toprule
        \textbf {Targets} & \textbf{\sys} & \textbf{\aflpp} & \textbf{\cmplog} & \textbf{\sysdict} & \textbf{\cmplogdict} \\ 
        \midrule
 bsdtar    &    \textbf{8,893} &    4,508 &      4,924 &      \textbf{9,194} &       7,811 \\
 exiv2     &    7,618 &    7,526 &      \textbf{9,213} &     10,631 &      \textbf{11,306} \\
 ffmpeg    &   \textbf{33,544} &   29,268 &     25,147 &     \textbf{33,821} &      22,692 \\
 jasper    &    3,811 &    3,400 &      \textbf{3,890} &      \textbf{4,257} &       3,870 \\
 nm-new    &    \textbf{5,475} &    4,049 &      5,150 &      \textbf{5,841} &       4,724 \\
 objdump   &    \textbf{7,196} &    5,484 &      6,508 &      \textbf{7,536} &       6,619 \\
 pdftotext &    \textbf{3,490} &    2,482 &      3,016 &      \textbf{3,410} &       2,982 \\
 readelf   &    \textbf{6,551} &    5,767 &      4,709 &      \textbf{6,462} &       5,549 \\
 size      &    \textbf{3,619} &    2,934 &      3,120 &      \textbf{4,037} &       3,167 \\
 strip-new &    \textbf{7,330} &    5,718 &      6,297 &      \textbf{7,930} &       6,306 \\
 tcpdump   &   16,996 &   13,399 &     \textbf{17,601} &     17,524 &      \textbf{17,841} \\
 tiff2pdf  &    \textbf{8,061} &    7,731 &      7,803 &      \textbf{7,974} &       7,830 \\
 tiff2ps   &    5,257 &    5,206 &      \textbf{5,411} &      5,306 &       \textbf{5,463} \\
 tiffcrop  &    \textbf{9,298} &    8,763 &      8,933 &      \textbf{9,254} &       8,835 \\
 xmllint   &    \textbf{8,347} &    5,179 &      5,392 &      8,628 &       \textbf{8,696} \\
        \midrule
        \multicolumn{2}{c|}{Mean gain} &  26.45\% & 16.98\%  &--- & 12.90\% \\
        \bottomrule
    \end{tabular}
\end{table}

\subsection{RQ2: Bug Discovery Effectiveness}\label{subsec:bugdiscovery}

In \textbf{RQ2}, we evaluate the bug discovery effectiveness of \sys compared to other state-of-the-art fuzzers. Following Klees et al.'s recommendation~\cite{klees2018evaluating} on using datasets with curated bugs, we tested our fuzzer on the Magma dataset~\cite{hazimeh2020magma}, comparing its performance with other fuzzers. Additionally, to understand \sys's real-world bug detection capabilities, we evaluate it on an array of widely-used standalone programs and libraries as specified in~\autoref{tab:programs}.

\vspace{0.1cm}
\noindent
{\bf Curated Bug Dataset.}
Magma, a popular dataset in the fuzzing community~\cite{lee2023learning,herrera2022dataflow}, contains 21 programs from nine open-source libraries with injected bugs. Our evaluation used 17 programs from eight of these libraries: \texttt{libpng}, \texttt{libtiff}, \texttt{libxml2}, \texttt{lua}, \texttt{openssl}, \texttt{poppler}, \texttt{sqlite3}, and \texttt{libsndfile}. \texttt{php} and its four fuzz drivers were the only targets omitted from our evaluation since this target encountered a compilation error during the final linking stage due to dependency incompatibility.  We compared \sys and \sysdict with fuzzers like AFL++, \cmplog, and \cmplogdict over five 24-hour test campaigns for each test program, totalling over 1 CPU-year of evaluation.

The results, detailed in \autoref{tab:magma}, reveal that \sys was the top performer in identifying unique bugs, outperforming or matching other fuzzers in all the evaluated programs. We observe that \sys's superior performance is consistent with Bohme et al.'s~\cite{bohme2022reliability} prior observation about a strong link between coverage and bug discovery. 

As an interesting side note, we observe that fewer bugs were triggered in \texttt{lua} and \texttt{openssl} compared to the other libraries. Upon investigation, we found this was due to a limited number of detectable ground-truth bugs in these programs. Specifically, \texttt{lua} had only four injected bugs~\cite{magma_lua}, and in \texttt{openssl}, only four have been confirmed to be reachable~\cite{hazimeh2020magma}, with several being provably unreachable~\cite{magma_openssl}.

\vspace{0.1cm}
\noindent
{\bf In-the-wild Bug Discovery.}
To evaluate \sys and \sysdict's bug discovery in real-world scenarios, we compared them with leading fuzzers on a dataset of common programs and libraries (\autoref{tab:programs}). After ten 24-hour campaigns per target, we analyzed each crash using standard deduplication practices ~\cite{klees2018evaluating}. Specifically, we employ stack hashes of the crashing inputs to perform deduplication and follow it up with manual analysis to validate deduplicated crashes are unique.

Our findings (\autoref{tab:wild_bugs}) show \sys and \cmplog discovering the same number of bugs, both surpassing AFL++. \sysdict, however, uncovered 20 unique bugs, including 2 Use After Frees (CWE-416), 3 Invalid Frees (CWE-761), 12 Assertion Violations (CWE-617), 1 Infinite Loop (CWE-835), and 2 Null Pointer Dereferences (CWE-476). Remarkably, 10 of these bugs were exclusively found by \sysdict, highlighting its superior bug discovery, especially given the heavily fuzzed nature of these targets.

Another interesting observation is that \sysdict heavily outperforms all the
other evaluation candidates, uncovering eleven more bugs than the next best
performing candidates (\sys, and \cmplog). A key contributor to this performance
difference is \texttt{xpdf} where \sysdict uncovered eight more bugs compared to
the other fuzzers. Looking closer at the coverage results, we see that \sysdict
uncovers on average $21.93\%$ more edges than the other fuzzers. In an effort to
understand if this coverage increase is correlated with the bugs uncovered by
\sysdict in \texttt{xpdf}, we measured Pearson's correlation coefficient between the edges
uncovered and unique crashes discovered across all the candidate fuzzers over the
ten 24-hour campaigns. We observed a strong correlation with a $0.88$ coefficient.
This observation is in line with previous findings of a strong correlation
between edges uncovered and bugs discovered while
fuzzing~\cite{bohme2022reliability}.

Finally, a testament to \sys's real-world impact is its identification of 12 bugs in \texttt{xpdf}, a popular PDF library, with eight being previously unknown as confirmed by the developers.

 \begin{table}[!h]
     \caption{\small\textbf{Cumulative number of unique bugs triggered and reached of \sys and \sysdict against three fuzzers in Magma programs for 24 hours over 5 runs. (triggered | reached)}}
     \centering
     \begin{adjustbox}{width=\columnwidth}
    \begin{tabular}{lrrr|rr}
     \toprule
         \textbf {Targets} & \textbf{\sys} & \textbf{\aflpp} & \textbf{\cmplog} & \textbf{\sysdict} & \textbf{\cmplogdict} \\ 
        \midrule
         tiffcp & 6 | 11 & 5 | 8 & 6 | 9 & 7 | 10 & 6 | 8 \\ 
         tiff\_read & 4 | 7 & 3 | 5 & 3 | 5 & 3 | 7 & 3 | 6\\ 
         libpng & 3 | 6 & 1 | 6 & 3 | 6 & 3 | 6 & 3 | 6\\ 
         xmllint & 3 | 7 & 2 | 7 & 3 | 7 & 3 | 8 & 3 | 8 \\ 
         libxml2\_xml & 5 | 9 & 3 | 8 & 3 | 8 & 4 | 9 & 4 | 9 \\ 
         lua & 2 | 4 & 1 | 2 & 1 | 2 & 1 | 2 & 1 | 2 \\ 
         asn1parse & 0 | 1 & 0 | 1 & 0 | 1 & 0 | 1 & 0 | 1 \\ 
         bignum & 0 | 1 & 0 | 1 & 0 | 1 & 0 | 1 & 0 | 1 \\ 
         asn1 & 2 | 4 & 2 | 4 & 2 | 4 & 2 | 4 & 2 | 4\\
         client & 1 | 7 & 1 | 7 & 1 | 7 & 1 | 5 & 1 | 7\\
         server & 2 | 6 & 1 | 6 & 1 | 6 & 1 | 4 & 1 | 6\\
         x509 & 1 | 5 & 0 | 5 & 0 | 5 & 0 | 5 & 0 | 5\\
         pdftoppm & 3 | 16 & 4 | 14 & 3 | 14 & 3 | 13 & 3 | 13\\
         pdf\_fuzzer & 3 | 16 & 3 | 13 & 2 | 13 & 2 | 8 & 2 | 12\\
         pdfimages & 4 | 13 & 5 | 10 & 4 | 11 & 3 | 11 & 4 | 11\\
         sqlite3\_fuzz & 3 | 10 & 5 | 13 & 3 | 11 & 3 | 10 & 4 | 13\\
         sndfile\_fuzzer & 7 | 8 & 7 | 8 & 7 | 8 & 8 | 8 & 7 | 8\\
         \midrule
         total & 49 | 131 & 43 | 118 & 42 | 118 & 44 | 112 & 44 | 120\\
         \bottomrule
     \end{tabular}
     \end{adjustbox}
     \label{tab:magma}
 \end{table}
 
\begin{table}[!ht]
\caption{\small\textbf{Cumulative number of unique bugs identified by \sys and \sysdict against three fuzzers in FuzzBench Programs and standalone programs for 24 hours over 10 runs}}
    \centering
    \setlength{\tabcolsep}{4pt}
    \renewcommand{\arraystretch}{1.1}
    \label{tab:wild_bugs}
    \begin{tabular}{lrrr|rr}
    \toprule
        \textbf {Targets} & \textbf{\sys} & \textbf{\aflpp} & \textbf{\cmplog} & \textbf{\sysdict} & \textbf{\cmplogdict} \\ 
        \midrule
  nm-new      &  1 &    0 &      1 &   2 &      1 \\
 size     & 2 &   2 &     1 &   2 &      1 \\
 objdump     &  1 &    0 &      1 &   1 &       1 \\
 pdftotext &  4 &    2 &      4 &    12 &       4 \\
 woff2     &  1 &    1 &      1 &    1 &       0 \\
 sqlite3    &  0 &    0 &      1 &    1 &       0 \\
 libxslt    &  0 &    0 &      0 &    1 &       0 \\
        \midrule
 total    &  9 &    5 &      9 &    20 &       7 \\
        \bottomrule
    \end{tabular}
\end{table}

\vspace{0.2cm}
\begin{longfbox}
\textbf{Result 2:} 
\sys triggers up to \maxmagma more ground-truth bugs compared to the state-of-the-art fuzzers on Magma and when coupled with dictionary uncovers 20 unique vulnerabilities in popular programs including eight that were previously unknown.
\end{longfbox}

\subsection{RQ3: Control Space Reduction}
\label{eval:controlspace}
In our stochastic control framework, the control space refers to the pool of seeds available for a fuzzer to select from during fuzzing. Unlike conventional coverage-guided fuzzers such as AFL++, which choose the next seed from a seed queue, \sys takes a different approach by scheduling based on frontier branches rather than individual seeds. AFL++, as the fuzzing campaign advances, adds more seeds to the queue, complicating the decision of which seed to schedule next. To address this complexity, AFL++ employs various techniques to manage the seed queue and prioritize specific seeds.

In contrast, \sys's scheduling strategy focuses on frontier branches. To assess the reduction in control space resulting from scheduling over frontier branches instead of seeds, we compare the number of frontier branches explored by \sys with the number of seeds explored by AFL++ throughout a fuzzing campaign. This comparison provides insights into the effectiveness of \sys's approach in streamlining the decision-making process during fuzzing.

\begin{figure}[!]
\centering
\includegraphics[scale=0.45]{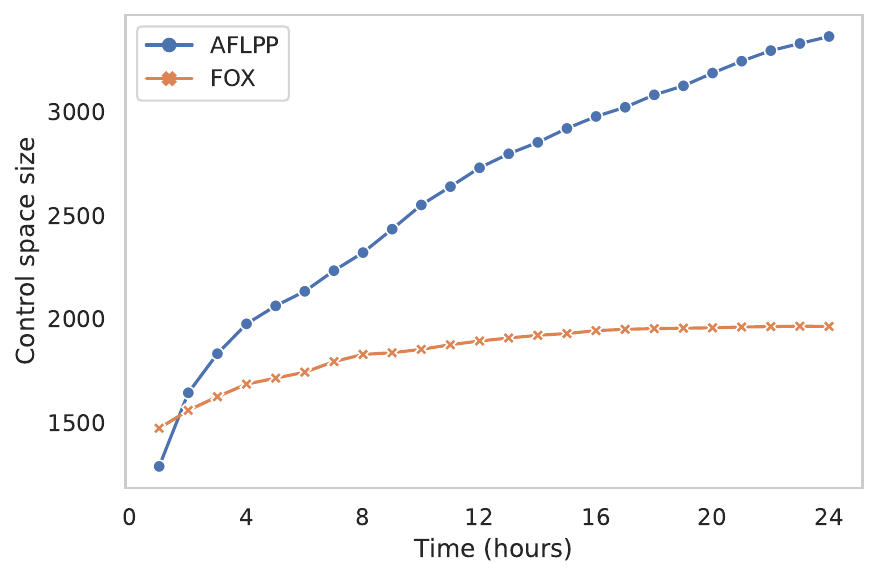}
\caption{\textbf{Control space comparison of \sys with \aflpp on \texttt{xmllint} over the course of a 24 hour fuzzing campaign.}}
\label{fig:controlspace}
\end{figure}

As illustrated in~\autoref{fig:controlspace}, the control space of AFL++ (i.e., the number of seeds) increases monotonically as it fuzzes \texttt{xmllint}, requiring more time to iterate through the entire seed corpus. In contrast, \sys reasons over a compact set of frontier branches that does not dramatically increase as the fuzzing campaign progresses. In fact, on some targets such as \texttt{readelf}, the set of frontier branches even decreases. The control space comparison between \sys and AFL++ on \numstandalone standalone programs at 12 and 24 hours, respectively, is presented in \autoref{tab:search}. On average, \sys demonstrates a remarkable reduction in the control space by $58.20\%$ during a 24-hour fuzzing campaign. Particularly noteworthy is \sys's ability to reduce the control space by a factor of 7 on programs like \texttt{readelf}.

\vspace{0.2cm}
\begin{longfbox}
\textbf{Result 3:} 
\sys reduces the control space by 49.18\% for 12-hour run and 58.20\% for 24-hour run when compared with AFL++.
\end{longfbox}

\begin{table}[!t]
    \caption{\small\textbf{Control space comparison of \sys with \aflpp on \numstandalone standalone programs at both 12 hours and 24 hours over 10 runs.}}
    \centering

   \begin{tabular}{lrr|rr}
   \toprule
    \multicolumn{1}{c}{\multirow{2}{*}{\textbf{Targets}}} & \multicolumn{2}{c}{\textbf{\# \sys Frontier}} & \multicolumn{2}{c}{\textbf{\# AFLPP Seeds}}  \\ \cline{2-5} 
\multicolumn{1}{c}{}& \multicolumn{1}{c}{\textbf {12 hrs}} & \textbf{24 hrs}  & \multicolumn{1}{c}{\textbf {12 hrs}} & \textbf{24 hrs}      \\
        \midrule
 bsdtar &        1,131 &        1,224 &          2,226 &          2,929 \\
 exiv2      &        2,953 &        3,057 &          2,121 &          2,446 \\
 ffmpeg     &        5,615 &        5,772 &          6,320 &          8,976 \\
 jasper     &        1,038 &        1,034 &          1,272 &          1,331 \\
 nm-new     &         971 &         944 &          1,970 &          2,772 \\
 objdump    &        1,299 &        1,285 &          3,239 &          3,940 \\
 pdftotext       &         929 &         981 &          1,466 &          1,661 \\
 readelf    &        1,561 &        1,215 &          7,075 &          8,970 \\
 size       &         671 &         670 &          1,470 &          1,692 \\
 strip-new  &        1,212 &        1,210 &          2,232 &          2,866 \\
 tcpdump    &        2,792 &        3,019 &          3,642 &          4,946 \\
 tiff2pdf   &        1,755 &        1,768 &          4,125 &          4,670 \\
 tiff2ps    &        1,093 &        1,098 &          2,316 &          2,692 \\
 tiffcrop   &        1,770 &        1,764 &          4,286 &          4,940 \\
 xmllint    &        1,894 &        1,964 &          2,731 &          3,366 \\
        \midrule
        \multicolumn{3}{c|}{Mean decrease} &  39.17 \% & 49.10\% \\
        \multicolumn{3}{c|}{Median decrease} & 49.18 \% & 58.20\% \\
        \bottomrule
    \end{tabular}
    \label{tab:search}
\end{table}
\subsection{RQ4: Ablation Study}
To quantify the contribution of the scheduling and mutation strategies employed by \sys~(\autoref{subsec:codecov}), we compare three different variants of \sys: (i) \sysbase: representing \sys with both the scheduling and mutator components deactivated and serving as the baseline; (ii) \sysched: employing only the optimization-guided scheduler; and (iii) \sys: enabling both the scheduler and the optimization-guided mutator. We measure the code coverage achieved by these three variants (\sysbase, \sysched, and \sys) on our \numstandalone standalone programs across 10 trials, each lasting 1 hour.

The results of our ablation study are presented in \autoref{table:ablation}. \sys achieves an average of $20.07\%$ more coverage than \sysbase and $9.23\%$ more coverage than \sysched. These results explicitly demonstrate the importance and contributions of both \sys scheduler and mutator in enhancing the overall performance. Notably, the frontier-branch-based scheduling alone results in an almost universal performance improvement of up to 27.2\% in mean coverage on \texttt{pdftotext} over the baseline, with the sole exception being \texttt{xmllint}. We attribute this anomaly to the fact that \texttt{xmllint} seems to have an unusually large control space where the number of frontier branches exceeds the number of seeds in the first hour. This trend is visualized in~\autoref{fig:controlspace}, comparing the control space of \sys against AFL++. We plan to further investigate this effect in the future. Similarly, enabling the Newton's method mutator yields a nearly unanimous gain of up to 31.9\% in mean coverage, with only a single exception of \texttt{tiff2ps}. We found that \texttt{tiff2ps} has a large proportion of non-convex branches. We discuss this phenomenon in greater depth in~\autoref{eval:introspection}. Improving our mutator for non-convex branches is an important direction that we plan to pursue in future work.

\vspace{0.2cm}
\begin{longfbox}
\textbf{Result 4:} 
Frontier-branch-based scheduling and Newton-step-based mutator both contribute to the 20.07\% edge coverage improvement of \sys over the baseline fuzzer in the first hour.
\end{longfbox}

\begin{table}[!ht]
\caption{\small\textbf{Mean edge coverage of \sys against its two variants on \numstandalone standalone programs over 10 fuzzing campaigns run for 1 hour each. We mark the highest number in bold.}  }
    \centering
    \label{table:ablation}
    \begin{tabular}{lrrr}
    \toprule
         \textbf {Targets} & \textbf{\sys} & \textbf{\sysbase} &\textbf{\sysched} \\ 
        \midrule
 bsdtar    &  \textbf{3,952} &       2,382 &        2,997 \\
 exiv2     &  \textbf{6,606} &       6,152 &        6,291 \\
 ffmpeg    & \textbf{17,316} &      16,291 &       17,272 \\
 jasper    &  \textbf{3,153} &       2,787 &        3,007 \\
 nm-new    &  \textbf{3,050} &       2,249 &        2,449 \\
 objdump   &  \textbf{4,996} &       3,672 &        4,017 \\
 pdftotext &  \textbf{2,567} &       1,905 &        2,423 \\
 readelf   &  \textbf{3,235} &       2,846 &        3,152 \\
 size      &  \textbf{2,621} &       2,131 &        2,351 \\
 strip-new &  \textbf{5,476} &       4,289 &        4,715 \\
 tcpdump   &  \textbf{5,875} &       4,656 &        5,523 \\
 tiff2pdf  &  \textbf{6,124} &       5,754 &        5,950 \\
 tiff2ps   &  4,373 &       4,202 &        \textbf{4,458} \\
 tiffcrop  &  \textbf{6,995} &       6,626 &        6,936 \\
 xmllint   &  4,220 &       \textbf{4,429} &        4,090 \\
        \midrule
        \multicolumn{2}{c|}{Mean gain} &  20.07\% & 9.23\%  \\
        \bottomrule
    \end{tabular}
\end{table}

 \begin{table}[!h]
     \caption{\small\textbf{Frontier branch flipping comparison between \sys, \cmplog, and \cmplogdict. R stands for frontier branches reached, and F stands for frontier branches flipped. We mark the highest number in bold.}}
     \centering
     \begin{adjustbox}{width=\columnwidth}
    \begin{tabular}{lrrrrrr|rrrr}
     \toprule
     \multicolumn{1}{c}{\multirow{2}{*}{\textbf{Targets}}} & \multicolumn{2}{c}{\textbf{\sys}} & \multicolumn{2}{c}{\textbf{AFLPP}} & \multicolumn{2}{c}{\textbf{\cmplog}}  &\multicolumn{2}{c}{\textbf{\sysdict}} &\multicolumn{2}{c}{\textbf{\cmplogdict}} \\ 
     \cline{2-11} 
 \multicolumn{1}{c}{}& \multicolumn{1}{c}{\textbf {R}} & \textbf{F} & \multicolumn{1}{c}{\textbf {R}} & \textbf{F} & \multicolumn{1}{c}{\textbf{R}} & \textbf{F} & \multicolumn{1}{c}{\textbf{R}} & \textbf{F} & \multicolumn{1}{c}{\textbf{R}} & \textbf{F}   \\
         \midrule
 bsdtar &  \textbf{2,873} & \textbf{1,694} & 1,714 & 776 & 1,837 & 890 & \textbf{3,083} &  \textbf{1,854} & 3,017 &  1,723 \\
 exiv2  &  3,929 &           948  & 3,718 & 886 & \textbf{4,477} & \textbf{1,088} & \textbf{5,664} &             \textbf{1,597} &                5,617 &                1,482 \\
 ffmpeg     &          \textbf{12,839} &           \textbf{7,800} &            10,219 &             6,093 &               8,922 &               4,478 &            \textbf{13,993} &             \textbf{7,766} &                8,908 &                3,921 \\
 jasper     &           \textbf{1,718} &            624 &             1,492 &              501 &               1,704 &                \textbf{631} &             \textbf{1,759} &              \textbf{638} &                1,696 &                 624 \\
 nm-new     &           \textbf{2,127} &      \textbf{1,153} &             1,591 &              802 &               1,964 &               1,056 &             \textbf{2,213} &             \textbf{1,158} &                1,840 &                 947 \\
 objdump    &           \textbf{2,746} &           \textbf{1,368} &             2,097 &             1,010 &               2,495 &               1,240 &             \textbf{2,817} &             \textbf{1,422} &                2,536 &                1,267 \\
 pdftotext       &           \textbf{1,425} &            \textbf{439} &             1,146 &              297 &               1,361 &                \textbf{439} &             \textbf{1,623} &             \textbf{574} &                1,342 &                 444 \\
 readelf    &           \textbf{1,475} &           \textbf{1,052} &             1,274 &              874 &               1,191 &                770 &             \textbf{1,706} &             \textbf{1,249} &                1,328 &                 893 \\
 size       &           \textbf{1,459} &            \textbf{771} &             1,228 &              619 &               1,281 &                671 &             \textbf{1,572} &              \textbf{850} &                1,303 &                 671 \\
 strip-new  &           \textbf{2,974} &           \textbf{1,526} &             2,421 &             1,212 &               2,615 &               1,381 &             \textbf{3,238} &             \textbf{1,633} &                2,625 &                1,351 \\
 tcpdump    &           6,601 &           \textbf{3,578} &             5,196 &             2,670 &               \textbf{6,789} &               3,496 &             6,816 &             \textbf{3,784} &                \textbf{6,850} &                3,592 \\
 tiff2pdf   &           \textbf{3,316} &           \textbf{1,575} &             3,211 &             1,483 &               3,227 &               1,487 &             \textbf{3,326} &             \textbf{1,572} &                3,230 &                1,502 \\
 tiff2ps    &           \textbf{2,155} &           \textbf{1,098} &             2,051 &              997 &               2,115 &               1,041 &             2,078 &             1,043 &                \textbf{2,133} &                \textbf{1,058} \\
 tiffcrop   &           \textbf{3,695} &           \textbf{1,887} &             3,590 &             1,784 &               3,642 &               1,812 &             \textbf{3,769} &             \textbf{1,942} &                3,611 &                1,784 \\
 xmllint    &           \textbf{3,502} &           \textbf{1,489} &             2,322 &              836 &               2,409 &                898 &             \textbf{3,894} &             \textbf{1,750} &                3,723 &                1,623 \\
         \midrule
         \multicolumn{3}{c|}{Mean gain} &  23.32\% & 33.99\% & 14.16\% & 21.05\% & ---& ---& 13.17\% & 19.47\% \\
         \bottomrule
     \end{tabular}
     \end{adjustbox}
     \label{tab:branchflip}
 \end{table}
\subsection{RQ5: \sys Performance Introspection}
\label{eval:introspection}
In \textbf{RQ5}, we delve into the sources of performance improvement of \sys and leverage the fine-grained information it provides about frontier branches to guide \sys mutator enhancements. Initially, we analyze how successful \sys is at reaching and flipping branches compared to other state-of-the-art fuzzers. Subsequently, we assess the relationship between branch convexity and \sys's ability to flip branches.

\vspace{0.1cm}
\noindent
{\bf Reaching and Flipping Branches.}
Distinguishing itself from traditional coverage-guided fuzzers, \sys focuses on targeting and flipping frontier branches to boost coverage. We expected \sys to not only reach but also flip more branches than its competitors. In 10x24-hour fuzzing campaigns, we compared the branch-flipping abilities of \sys and \sysdict with AFL++, \cmplog, and \cmplogdict (\autoref{tab:branchflip}). The results show that both \sys and \sysdict reach and successfully flip more frontier branches, with \sys flipping 33.99\% more than AFL++ and up to 21.05\% more than \cmplog, while \sysdict surpasses \cmplogdict by 19.47\%. This underscores \sys's effective framework in achieving new coverage by handling different types of frontier branches.

\vspace{0.1cm}
\noindent
{\bf Branch Convexity Estimation.}
As \sys's Newton's method mutator is primarily optimized for convex functions~\cite{Boyd_Vandenberghe_2004}, we test \sys on \numstandalone standalone programs over 5x24-hour campaigns to evaluate its performance on both convex and non-convex branches. Identifying branch convexity is challenging, so we used midpoint convexity checks during fuzzing as an estimate. We consider two inputs to be midpoint convex if the branch distance of the average of the two inputs is less than or equal to the average of the branch distances of the two inputs.

The experiment involved tracking midpoint convexity before and after applying Newton's method. For each branch, we calculated a ratio of successful midpoint checks to branch reaches, indicating convexity. Higher ratios suggest convex behavior, while lower ratios indicate non-convexity. We expected \sys to excel in flipping convex branches. This was analyzed using logistic regression, modeling branch flips against the convexity estimate and its interaction with the binary.

The regression (details in \autoref{eval: logit}) showed a positive correlation between branch convexity and \sys's flipping success, with a McFadden's pseudo-R-squared of 0.048, indicating a good fit~\cite{domencich_urban_1975, hensher2021behavioural}. The correlation between branch convexity and flipping success varied among programs, which was anticipated. Although midpoint convexity is a useful estimate, it does not always precisely represent a branch's true convexity and feasibility. For instance, non-convex branches might pass the convexity check in certain regions, whereas some convex branches could be infeasible, i.e., impossible to flip.

\vspace{0.2cm}
\begin{longfbox}
\textbf{Result 5:} 
\sys can flip up to $33.99\%$ more frontier branches, showcasing the efficacy of its stochastic control-guided framework.
\end{longfbox}

\section{Limitations and Future Work}
\label{sec:limitations}
In this section, we discuss the theoretical and engineering limitations of the current design of \sys, and outline potential future directions for improvement.

\noindent\textbf{Newton's Method.} 
\sys utilizes Newton's method to generate inputs that decrease branch distances and potentially flip frontier branches. The main advantage of this approach is its simplicity and efficiency — it only needs two distinct inputs that reach the target branch to estimate a Newton step. However, Newton's method struggles with non-convex functions like checksums, hash functions, and complex string operations. We aim to enhance our approach by selectively applying more accurate, albeit computationally intensive methods for hard-to-flip non-convex branches.

\noindent\textbf{Local Search.}
The local search module of \sys establishes a local linear approximation (w.r.t. a seed) of branch distance functions, utilized by Newton's method for generating new inputs. Despite the accuracy demonstrated by methods like REDQUEEN \cite{Aschermann2019}, their complexity and inefficiency make them less practical compared to local search, as evidenced in our experimental findings. Currently, local search utilizes AFL++ \texttt{havoc} for mutations, which has difficulty in adaptively controlling the extent and scale of mutations, resulting in inefficient and redundant mutations. In the future, we aim to enhance \sys by integrating a sophisticated mutator capable of dynamically adjusting its mutation strategy based on past successes in reaching the frontier branches.

\noindent\textbf{Frontier Branch Scheduler.}
Our scheduler selects seeds associated with promising frontier branches, refining the control space more effectively than traditional seed-based methods. However, the scheduler depends on estimates of the likelihood of reducing branch distance to identify promising branches. Despite the theoretical optimality demonstrated in~\autoref{thm proof: greedy schedule optimal}, its practical efficacy relies on the accuracy of these estimates. In fact, we have identified several pathological branches within target programs where the mutator consistently decreases branch distances but fails to flip them throughout the entire fuzzing campaign. We speculate that such behavior suggests the infeasibility of these branches. Therefore, we plan to modify the scheduler in the future to deploy an aggressive re-weighting strategy to detect and deprioritize such pathological branches.

\noindent\textbf{ Engineering Limitations.} The current prototype of \sys is optimized for integer and string comparisons, which are prevalent in target applications, as shown by our coverage (\autoref{subsec:codecov}) and bug discovery (\autoref{subsec:bugdiscovery}) performance. However, our prototype currently does not support branches with floating point operations or variable-variable string comparisons, which appear in benchmarks like \texttt{lcms} and \texttt{proj4}. This limitation occasionally results in lower coverage compared to leading fuzzers. Future updates will aim to expand \sys's applicability to these branch types and involve the community in enhancing the tool through open-source contributions.

\section{Related Work}

\noindent\textbf{Search-based Software Testing.}
Such approaches use metaheuristic optimization for tasks like test-case generation~\cite{mcminn2011search,lakhotia2010austin,fraser2011evosuite}. Szekeres et al.~\cite{szekeressearch} suggested using stochastic local search and taint tracking for targeted fuzzing of specific branches to minimize the distance to a target branch. However, \sys differs by using local search to estimate the gradient of the branch distance function, employing Newton's method for more effective root finding. There are also hybrid methods that integrate path constraint information into their search strategy~\cite{choi2019grey,cadar2008klee,stephens2016driller,godefroid2008automated}, but these face scaling challenges in large programs. \sys avoids these challenges by focusing on frontier branches and using branch distance as a metric to create inputs that flip these branches.

\noindent{\textbf{Mutator Policies.}}
Several prior studies have aimed to optimize mutation policies for fuzzing~\cite{chen2018angora, she2019neuzz, lyu2019mopt,Aschermann2019, lee2023learning}. REDQUEEN~\cite{Aschermann2019} employed input colorization with strong assumptions about input consistency, contrasted by \sys's flexible mutations. Angora~\cite{chen2018angora} used gradient descent and taint tracking, while \sys identifies input bytes using local search and Newton's method. Neuzz~\cite{she2019neuzz} adopted a neural network approach for branch behavior, whereas \sys constructs a simpler model using fewer samples. Lastly, FairFuzz~\cite{lemieux2018fairfuzz} focused on rare branches, whereas \sys uses a branch distance metric for targeted branch-flipping inputs.

\noindent{\textbf{Scheduler Policies.}}
K-Scheduler~\cite{She2022EffectiveSS} uses a centrality-based seed scheduling approach that might overestimate the potential of seeds to discover new edges, possibly including many unreachable nodes in the entire CFG. In contrast, \sys uses data solely from frontier branches for scheduling and mutation, leading to more accurate decisions.

\noindent{\textbf{Feedback Metrics.}}
Several studies use detailed feedback like path coverage for seed selection~\cite{yue2020ecofuzz,bohme2016coverage}, which can lead to seed explosion~\cite{wang2019sensitive}. \sys uses frontier branch granularity and branch distances to manage this issue. Other research proposes different feedback metrics, like data-flow coverage~\cite{herrera2022dataflow,gan2020greyone} or objectives like reaching sanitizer instrumentation~\cite{osterlund2020parmesan,chen2020savior}, to enhance bug discovery by exploring various aspects of program state space. In the future, we aim to investigate how \sys can be combined with these diverse fuzzing objectives.

\section{Conclusion}
In conclusion, this paper presents a unified framework for coverage-guided mutation-based fuzzing by treating fuzzing as an online stochastic control problem. Our extensive experimental results demonstrate that our proof-of-concept implementation \sys can address many challenges of coverage-guided fuzzing for large and complex programs.

\section*{Acknowledgements}

We thank our shepherd and the anonymous reviewers for their constructive and
valuable feedback. We would also like to thank László Szekeres, Jonathan
Metzman, Franjo Ivancic, Xinhao Yuan, Junfeng Yang, and Baishakhi Ray for their
helpful comments and discussions that improved the paper. This work is supported
partially by an NSF CAREER award; a Google Faculty Fellowship; and an award from
the Google Cyber NYC Institutional program. Any opinions, findings, conclusions,
or recommendations expressed herein are those of the authors, and do not
necessarily reflect those of NSF, or Google.

\bibliographystyle{plain}
\bibliography{paper}

\appendices
\section{Theory Addendum}
\subsection{Proof of Theorem \ref{thm: greedy schedule optimal} \label{thm proof: greedy schedule optimal}}
\begin{proof}
Let expected new coverage gain given seed list $S_i$ and chosen seed index $u_i$ be defined as:
\begin{equation}
\sum_{b \in B_i} \Pr(b \text{ flips} \mid mut(S_i[u_i]))
\label{eq: thm1 orig sum}
\end{equation}
We can transform~\autoref{eq: thm1 orig sum} to sum over the set of all branches $B$ instead of the set of not-flipped branches at stage $i$ $B_i$:
\begin{multline}
\sum_{b \in B} \Pr(b \text{ not flipped prior } i) \Pr(b \text{ flips} \mid mut(S_i[u_i]))
\label{eq: thm1 new sum}
\end{multline}
Now consider an optimal schedule $\pi = [u_1, u_2, ..., u_n]$. Given $\pi$, we can obtain another assignment $\pi'$ from $\pi$ by exchanging $u_1$ with a $u'_1$ such that $u'_1$ is the first greedy choice:
\begin{multline}
\sum_{b \in B} \Pr(b \text{ flips} \mid mut(S_1[u_1])) \\ \leq \sum_{b \in B} \Pr(b \text{ flips} \mid mut(S_1[u'_1]))
\end{multline}
To simplify the equations below, let:
\begin{equation}
\begin{gathered}
p = \Pr(b \text{ flips} \mid mut(S_1[u_1])) \\
p' = \Pr(b \text{ flips} \mid mut(S_1[u'_1])) \\
q = \max_i \Pr(b \text { flips} \mid mut(S_i[u_i]))
\end{gathered}
\end{equation}
The total expected coverage gain for $u_1$ is then:
\begin{equation}
    \sum_{b \in B}\left[p + (1-p)(q + (1-q)q + ... + (1-q)^{n-2}q)\right]
\label{eq: thm1 exp u_1}
\end{equation}
and analogously for $u'_1$:
\begin{equation}
    \sum_{b \in B}\left[p' + (1-p')(q + (1-q)q + ... + (1-q)^{n-2}q)\right]
\label{eq: thm1 exp u'_1}
\end{equation}
Subtracting~\autoref{eq: thm1 exp u_1} from~\autoref{eq: thm1 exp u'_1}, we get:
\begin{equation}
\sum_{b \in B}\left[(p'-p)(1 - (q + (1-q)q + ... + (1-q)^{n-2}q))\right]
\end{equation}
Note that the term involving $q$ is in fact a CDF of the geometric distribution. We can therefore simplify further:
\begin{equation}
\sum_{b \in B}\left[(p'-p)(1 - (1-(1-q)^{n-1}))\right]
\end{equation}
\begin{equation}
\sum_{b \in B}\left[(p'-p)(1-q)^{n-1}\right]
\end{equation}
Since $q$ is a probability, therefore $0 \leq q \leq 1$, the equation above can be bounded as follows:
\begin{equation}
    0 \leq \sum_{b \in B}\left[(p'-p)(1-q)^{n-1}\right] \leq \sum_{b \in B}(p'-p)
\end{equation}
Note that $p'-p \geq 0$ since $u'_1$ is the first greedy choice. Therefore, the total expected coverage gain of $u'_1$ is at least as high as the total expected coverage gain of the optimal assignment $u_1$. This argument exactly applies to the assignment $u_2, u_3, ... ,u_n$. Therefore, the greedy scheduling strategy is the optimal scheduling strategy for fuzzing as online stochastic control.
\end{proof}

\section{Performance Characteristics}

\subsection{Mann-Whitney U Test \label{eval:mann-whitney-u}}

For the code coverage evaluation done as part
of answering \textbf{RQ1} (\autoref{subsec:codecov}), we report the results of significance
testing by performing the Mann-Whitney U test~(as recommended by Klees
et.al~\cite{klees2018evaluating}) to validate that the performance improvement
\sys and \sysdict obtain over the other state-of-the-art fuzzers is
statistically significant and not due to the randomness of the fuzzing process.

The results of the Mann-Whitney U test performed over
the coverage results obtained for standalone programs and the FuzzBench programs
are presented in~\autoref{eval:standalone_sign}
and~\autoref{eval:fuzzbench_sign}, respectively. As evident from~\autoref{eval:standalone}, there is a large, statistically significant performance improvement obtained by \sys compared to AFL++ in all standalone
targets but \texttt{exiv2} and \texttt{tiff2ps}. When comparing \sys against \cmplog, the only
instances where the performance difference is not statistically significant is \texttt{jasper} and \texttt{tcpdump}.
Interestingly, \texttt{jasper} and \texttt{tcpdump} are the two of the four targets where \cmplog
appears to have a slight performance advantage over \sysdict, but as shown with Mann-Whitney this improvement is statistically insignificant. Similarly, \texttt{tcpdump} and \texttt{xmllint} are the two out of the four targets where \cmplogdict seems to perform better than \sysdict, yet this advantage is also statistically insignificant. In the case of the FuzzBench dataset (\autoref{eval:fuzzbench}), the only four targets where AFL++ appears to outperform \sys — \texttt{harfbuzz}, \texttt{libjpeg-turbo}, \texttt{proj4}, and \texttt{re2} — are statistically insignificant. Similarly, \cmplog only statistically significantly outperforms \sys on 4 targets out of the 10 targets \cmplog has a higher mean coverage on compared to \sys. Likewise, \cmplogdict's better coverage performance on \texttt{bloaty}, \texttt{systemd}, and \texttt{vorbis} is statistically insignificant.

\begin{table}[!h]
\caption{\small\textbf{Mann-Whitney U test results of \sys against \aflpp, \cmplog, and \sysdict against \cmplogdict, on \numstandalone standalone programs for 24 hours over 10 runs. Statistically significant results ($p < 0.05$) are in bold.}}
    \label{eval:standalone_sign}
    \begin{tabular}{lrrr}
        \toprule
        \textbf {Targets} & \textbf{\aflpp} & \textbf{\cmplog} & \textbf{\cmplogdict} \\ 
        \midrule
 bsdtar    & \textbf{1.83e-04} &  \textbf{1.83e-04} &   \textbf{1.00e-03} \\
 exiv2     & 1.86e-01 &  \textbf{5.83e-04} &   \textbf{1.73e-02} \\
 ffmpeg    & \textbf{5.80e-03} &  \textbf{3.30e-04} &   \textbf{1.83e-04} \\
 jasper    & \textbf{2.46e-04} &  3.07e-01 &   2.41e-01 \\
 nm-new    & \textbf{1.83e-04} &  \textbf{4.59e-03} &   \textbf{1.82e-04} \\
 objdump   & \textbf{1.82e-04} &  \textbf{3.26e-04} &   \textbf{1.83e-04} \\
 pdftotext & \textbf{1.82e-04} &  \textbf{2.20e-03} &   7.34e-01 \\
 readelf   & \textbf{1.01e-03} &  \textbf{5.83e-04} &   1.62e-01 \\
 size      & \textbf{1.82e-04} &  \textbf{7.69e-04} &   \textbf{1.83e-04} \\
 strip-new & \textbf{1.83e-04} &  \textbf{3.30e-04} &   \textbf{1.83e-04} \\
 tcpdump   & \textbf{1.83e-04} &  6.40e-02 &   3.07e-01 \\
 tiff2pdf  & \textbf{1.71e-03} &  \textbf{7.28e-03} &   \textbf{7.28e-03} \\
 tiff2ps   & 2.73e-01 &  \textbf{1.31e-03} &   \textbf{1.71e-03} \\
 tiffcrop  & \textbf{1.83e-04} &  \textbf{2.83e-03} &   \textbf{1.01e-03} \\
 xmllint   & \textbf{1.83e-04} &  \textbf{1.83e-04} &   7.91e-01 \\
        \bottomrule
    \end{tabular} 
\end{table}

\begin{table}[!h]
\caption{\small\textbf{Mann-Whitney U test results of \sys against \aflpp, \cmplog and \sysdict against \cmplogdict on \numfuzzbench FuzzBench programs for 24 hours over 10 runs. Statistically significant results ($p < 0.05$) are in bold.}}
    \label{eval:fuzzbench_sign}
    \begin{tabular}{lrrr}
        \toprule
        \textbf {Targets} & \textbf{\aflpp} &\textbf{\cmplog} & \textbf{\cmplogdict} \\ 
        \midrule
 bloaty        & 5.38e-02 &  5.71e-01 &   6.40e-02 \\
 curl          & 3.45e-01 &  7.57e-02 &   \textbf{3.19e-03} \\
 freetype      & \textbf{1.83e-04} &  \textbf{1.83e-04} &   \textbf{1.83e-04} \\
 harfbuzz      & 6.23e-01 &  \textbf{1.83e-04} &   \textbf{1.83e-04} \\
 jsoncpp       & 9.30e-01 &  7.74e-02 &   \textbf{1.59e-05} \\
 lcms          & \textbf{1.53e-04} &  \textbf{6.63e-04} &   1.00e+00 \\
 libjpeg-turbo & 9.39e-01 &  1.00e+00 &   \textbf{3.09e-02} \\
 libpcap       & \textbf{6.39e-05} &  7.57e-02 &   \textbf{3.11e-02} \\
 libpng        & \textbf{1.67e-04} &  5.18e-01 &   \textbf{2.29e-02} \\
 libxml2       & \textbf{4.40e-04} &  \textbf{2.83e-03} &   \textbf{1.82e-04} \\
 libxslt       & \textbf{1.70e-03} &  \textbf{2.32e-02} &   \textbf{1.13e-02} \\
 mbedtls       & \textbf{1.82e-04} &  \textbf{1.70e-03} &   \textbf{2.83e-03} \\
 openh264      & 2.41e-01 &  \textbf{1.70e-03} &   \textbf{4.59e-03} \\
 openssl       & \textbf{1.76e-04} &  \textbf{1.78e-04} &   \textbf{1.77e-04} \\
 openthread    & \textbf{3.40e-02} &  \textbf{1.71e-03} &   \textbf{3.60e-03} \\
 proj4         & 5.21e-01 &  \textbf{1.83e-04} &   2.41e-01 \\
 re2           & 1.60e-01 &  8.20e-01 &   \textbf{1.12e-02} \\
 sqlite3       & 9.70e-01 &  6.23e-01 &   \textbf{2.57e-02} \\
 stb           & \textbf{2.08e-04} &  5.37e-02 &   \textbf{3.17e-03} \\
 systemd       & \textbf{4.40e-04} &  4.73e-01 &   2.41e-01 \\
 vorbis        & 1.84e-01 &  2.26e-01 &   5.43e-01 \\
 woff2         & \textbf{1.81e-04} &  \textbf{1.31e-03} &   \textbf{3.30e-04} \\
 zlib          & 5.01e-01 &  \textbf{2.43e-03} &   \textbf{2.05e-04} \\
        \bottomrule
    \end{tabular}
\end{table}

\subsection{Throughput Measurements \label{eval:throughput}}

We report the mean number of mutations (in millions) exercised by \sys and \sysdict compared to AFL++, \cmplog, and \cmplogdict as part
of our coverage evaluation and report them for the Fuzzbench programs
in~\autoref{app:fuzzbench_throughput} and the standalone programs
in~\autoref{app:standalone_throughput}. In both instances, we see that both \sys
and \sysdict are largely comparable in performance to \cmplog and \cmplogdict,
showcasing the lightweight nature of our stochastic-control-guided approach. 

An interesting insight is that in certain targets \sys achieves higher
throughput than AFL++. This occurs because of two reasons. First, the scheduler
takes into account the productive time spent on a frontier branch compared to
the total time spent on it, thereby indirectly penalizing low throughput seeds.
Second, FOX dynamically turns off instrumentation for flipped (i.e., already
reached) branches.

\begin{table}[!h]\caption{\small\textbf{Mean number of executions (in millions) of \sys and \sysdict against four fuzzers on \numfuzzbench FuzzBench programs for 24 hours over 10 runs.}  }
    \centering
    \setlength{\tabcolsep}{2pt}
    \renewcommand{\arraystretch}{1.1}
    \label{app:fuzzbench_throughput}
    \begin{tabular}{lrrr|rr}
\toprule
\textbf{Targets}    &     \textbf{\sys} &     \textbf{\aflpp} &   \textbf{\cmplog} &    \textbf{\sysdict} &   \textbf{\cmplogdict} \\
\midrule
 bloaty        &   18.63 &   12.15 &     14.25 &      9.96 &      17.32 \\
 curl          &  212.34 &  388.11 &    400.29 &    201.40 &     399.97 \\
 freetype      &  133.77 &  139.14 &    207.23 &    183.36 &     167.00 \\
 harfbuzz      &  102.33 &  107.82 &     10.03 &    104.95 &      28.09 \\
 jsoncpp       &  320.80 &  288.62 &    315.40 &    583.72 &     459.06 \\
 lcms          &  200.69 &  770.22 &    664.99 &    196.91 &     579.56 \\
 libjpeg-turbo &  257.84 &  470.42 &    584.65 &    281.14 &     433.42 \\
 libpcap       &    5.91 &   17.17 &      8.85 &      5.79 &      14.19 \\
 libpng        &  419.08 &  469.73 &    474.01 &    414.27 &     340.42 \\
 libxml2       &   61.89 &   40.95 &     47.00 &     62.06 &      40.04 \\
 libxslt       &  585.16 &  600.73 &    641.32 &    541.56 &     622.92 \\
 mbedtls       &  205.98 &  230.64 &    289.20 &    119.89 &     379.54 \\
 openh264      &   10.66 &    9.77 &      8.86 &     16.99 &      12.62 \\
 openssl       &  249.52 &  650.18 &    591.01 &    220.17 &     728.32 \\
 openthread    &  123.44 &   85.86 &     83.29 &    123.97 &      78.14 \\
 proj4         &   52.22 &  108.53 &     86.30 &     56.87 &     104.03 \\
 re2           &  170.98 &  358.33 &    378.28 &    169.59 &     364.55 \\
 sqlite3       &   58.67 &   70.83 &     90.31 &     84.01 &      99.10 \\
 stb           &  148.54 &  252.21 &    243.58 &    173.19 &     271.04 \\
 systemd       &    8.25 &    8.48 &      8.54 &      8.70 &       8.40 \\
 vorbis        &  441.68 &  322.17 &    185.16 &    413.96 &     173.54 \\
 woff2         &  299.36 &  915.85 &    115.84 &    306.54 &     118.81 \\
 zlib          &   16.89 &  686.02 &    649.81 &     17.24 &     566.66 \\
\bottomrule
\end{tabular}
\end{table}

\begin{table}[!ht]\caption{\small\textbf{Mean number of executions (in millions) of \sys and \sysdict against four fuzzers on \numstandalone standalone programs for 24 hours over 10 runs.}  }
    \centering
    \setlength{\tabcolsep}{3pt}
    \renewcommand{\arraystretch}{1.1}
    \label{app:standalone_throughput}
    \begin{tabular}{lrrr|rr}
\toprule
\textbf{Targets}    &     \textbf{\sys} &     \textbf{\aflpp} &   \textbf{\cmplog} &    \textbf{\sysdict} &   \textbf{\cmplogdict} \\
\midrule
 bsdtar    &   15.25 &    7.93 &      8.95 &     14.85 &      10.43 \\
 exiv2     &    6.13 &    8.25 &      7.64 &      5.54 &       4.52 \\
 ffmpeg    &    3.84 &    3.14 &      3.11 &      3.89 &       2.52 \\
 jasper    &   14.40 &   14.06 &     14.20 &     18.14 &      13.39 \\
 nm-new    &    8.37 &    9.95 &     13.02 &      9.12 &       8.94 \\
 objdump   &    7.68 &    7.90 &      9.22 &      8.50 &       8.84 \\
 pdftotext &    4.67 &    4.09 &      4.05 &      4.74 &       4.37 \\
 readelf   &   13.19 &    9.64 &      9.53 &     10.42 &      11.52 \\
 size      &   10.26 &   13.85 &     12.43 &      7.52 &       8.91 \\
 strip-new &   12.34 &    9.27 &     11.72 &     12.50 &       9.53 \\
 tcpdump   &    9.08 &    9.53 &      9.69 &      8.79 &      11.27 \\
 tiff2pdf  &   11.74 &   10.73 &     10.87 &     12.65 &      10.84 \\
 tiff2ps   &   13.85 &   11.43 &     12.27 &     18.63 &      21.42 \\
 tiffcrop  &   16.51 &   10.23 &     12.32 &     14.34 &       9.69 \\
 xmllint   &    8.75 &    7.44 &      9.50 &     13.79 &       9.39 \\
\bottomrule
\end{tabular}
\end{table}

\pagebreak

\subsection{Types of Bugs Identified by \sysdict \label{eval:bugtype}}

\autoref{tab:bugtype} specifies the bugs found by \sysdict as part of the bug
discovery evaluation performed on the evaluation dataset consisting of both
standalone programs and library drivers.

\begin{table}[h!]\caption{\small\textbf{The types of bugs found by \sysdict}  }
    \label{tab:bugtype}
    \centering
    \setlength{\tabcolsep}{3pt}
    \renewcommand{\arraystretch}{1.1}
    \begin{tabular}{lr}
\toprule
\textbf{Targets}    &     \textbf{Bug Type} \\
\midrule
nm-new & Use After Free \\
nm-new & Invalid Free \\ 
size & Use After Free \\ 
size & Invalid Free \\ 
objdump & Invalid Free \\ 
xpdf & Assertion Violation \\ 
xpdf & Assertion Violation \\
xpdf & Assertion Violation \\ 
xpdf & Assertion Violation \\
xpdf & Assertion Violation \\ 
xpdf & Assertion Violation \\ 
xpdf & Assertion Violation \\
xpdf & Assertion Violation \\
xpdf & Assertion Violation \\ 
xpdf & Assertion Violation \\
xpdf & Assertion Violation \\
xpdf & Stack Overflow \\
woff2 & Null Pointer Dereference \\
sqlite3 & Assertion Violation \\
libxslt & Null Pointer Dereference\\ 
\bottomrule
\end{tabular}
\end{table}

\subsection{Coverage-over-time\label{app:coverage}}

\begin{figure*}[ht!]
        \includegraphics[width=\linewidth]{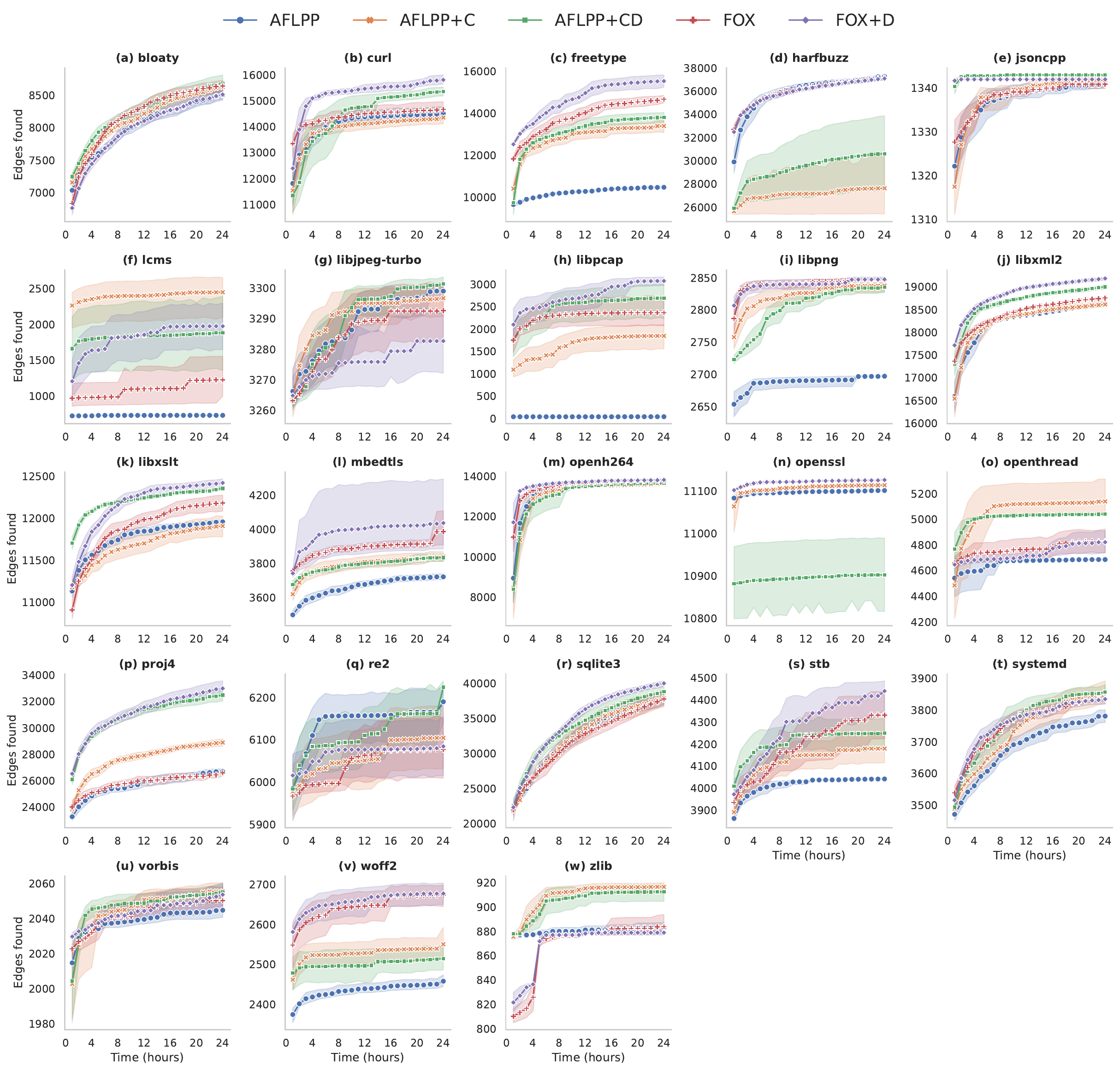}
	\caption{\small\textbf{The arithmetic mean edge coverage for \sys and \sysdict against three other fuzzers running for 24 hours over 10 runs on the FuzzBench programs. The error bars indicate one standard deviation.}}
	\label{fig:coverage}
\end{figure*}

We compare the code coverage performance of \sys and \sysdict against the other
state-of-the-art fuzzers (AFL++, \cmplog, and \cmplogdict) on the FuzzBench
dataset running 10 campaigns of 24 hours each. \autoref{fig:coverage} showcases
the mean edge coverage achieved by the fuzzers across the 10 runs over time with
the error bars indicating one standard deviation.

\pagebreak
\section{Performance Introspection Details}

\label{eval: logit}

\subsection{Logistic Regression Result}

\begin{table*}[h!]
\centering
\caption{\textbf{Logistic regression model of \sys flipping a branch on \numstandalone standalone programs over 5 fuzzing campaign runs 24 hour each. The table shows fitted parameters for the model: \textit{branch flipped $\sim$ convexity + target + convexity $\times$ target}.}}
\label{tab: logit}
\setlength{\tabcolsep}{2pt} %
\begin{tabular}{lrrrrr}
\toprule
\textbf{Variable} & \textbf{Coeff} & \textbf{SE} & \textbf{p-val} & \multicolumn{2}{c}{\textbf{95 \% CI}} \\
\midrule
 intercept                   &        -1.461 &            0.122 &     0.000 &   -1.699 &    -1.222 \\
 ffmpeg            &         0.296 &            0.129 &     0.021 &    0.045 &     0.548 \\
 jasper            &         0.228 &            0.150 &     0.128 &   -0.065 &     0.521 \\
 libarchive        &         1.387 &            0.140 &     0.000 &    1.113 &     1.661 \\
 nm-new            &         0.324 &            0.154 &     0.035 &    0.023 &     0.625 \\
 objdump           &         0.474 &            0.169 &     0.005 &    0.143 &     0.805 \\
 size              &         0.672 &            0.182 &     0.000 &    0.315 &     1.028 \\
 strip-new         &         0.335 &            0.152 &     0.028 &    0.036 &     0.633 \\
 tcpdump           &         0.981 &            0.126 &     0.000 &    0.734 &     1.229 \\
 tiff2pdf          &         0.090 &            0.144 &     0.533 &   -0.193 &     0.373 \\
 tiff2ps           &         0.359 &            0.150 &     0.017 &    0.066 &     0.653 \\
 tiffcrop          &        -0.023 &            0.142 &     0.870 &   -0.301 &     0.255 \\
 xmllint           &         1.022 &            0.146 &     0.000 &    0.735 &     1.309 \\
 xpdf              &         1.122 &            0.232 &     0.000 &    0.668 &     1.576 \\
 convexity                      &         1.013 &            0.220 &     0.000 &    0.582 &     1.444 \\
 convexity $\times$ ffmpeg     &        -0.408 &            0.234 &     0.081 &   -0.866 &     0.050 \\
 convexity $\times$ jasper     &        -1.040 &            0.304 &     0.001 &   -1.636 &    -0.444 \\
 convexity $\times $ libarchive &        -0.739 &            0.261 &     0.005 &   -1.252 &    -0.227 \\
 convexity $\times $ nm-new     &         0.370 &            0.267 &     0.165 &   -0.153 &     0.893 \\
 convexity $\times $ objdump    &         0.594 &            0.292 &     0.042 &    0.022 &     1.167 \\
 convexity $\times $ size       &         0.810 &            0.313 &     0.010 &    0.196 &     1.424 \\
 convexity $\times $ strip-new  &         0.931 &            0.274 &     0.001 &    0.394 &     1.469 \\
 convexity $\times $ tcpdump    &        -0.211 &            0.237 &     0.373 &   -0.676 &     0.254 \\
 convexity $\times $ tiff2pdf   &         0.400 &            0.276 &     0.147 &   -0.141 &     0.942 \\
 convexity $\times $ tiff2ps    &         0.194 &            0.286 &     0.497 &   -0.366 &     0.754 \\
 convexity $\times $ tiffcrop   &         0.347 &            0.272 &     0.201 &   -0.185 &     0.880 \\
 convexity $\times $ xmllint    &         0.087 &            0.271 &     0.750 &   -0.445 &     0.618 \\
 convexity $\times $ xpdf       &        -1.719 &            0.458 &     0.000 &   -2.617 &    -0.822 \\
\bottomrule
\end{tabular}
\end{table*}

We evaluate if there is a correlation between the degree to which we estimate a
branch to be convex and the ability of FOX to flip the branch in consideration.
We perform this analysis to support our claim that our Newton-based mutator can
flip branches behaving as convex functions. \autoref{tab: logit} shows the
result of the logistic regression analysis as discussed in
\autoref{eval:introspection}. We include the coefficient (\textbf{Coeff}), the
standard error (\textbf{SE}), two-tailed p-value for the t-statistic
(\textbf{p-val}), and the 95\% confidence intervals (\textbf{95\% CI}).

\end{document}